\documentclass[twocolumn,reprint,superscriptaddress,amsmath,amssymb,aps,nofootinbib]{revtex4-1}
\pdfoutput=1

\usepackage{multirow}
\usepackage{graphicx}
\usepackage{amsmath}
\usepackage{amssymb}
\usepackage{amsfonts}
\usepackage{physics}
\usepackage[dvipsnames]{xcolor}
\usepackage[linktoc=none]{hyperref}
\usepackage[utf8]{inputenc}
\usepackage{braket}
\usepackage{float}
\usepackage{tikz}
\usepackage{diagbox}

\hypersetup{colorlinks,linkcolor={blue},citecolor={teal},urlcolor={violet}}

\newcommand{\documenttitle}{Constraining scalar leptoquarks using COHERENT data}

\newcommand{\cns}{CE$\nu$NS }
\newcommand{\INFN}{INFN - Sezione di Napoli, Complesso Univ. Monte S. Angelo, I-80126 Napoli, Italy}
\newcommand{\UNINA}{Dipartimento di Fisica ``Ettore Pancini'', Università degli studi di Napoli ``Federico II'', Complesso Univ. Monte S. Angelo, I-80126 Napoli, Italy}
\newcommand{\SSM}{Scuola Superiore Meridionale, Università degli studi di Napoli ``Federico II'', Largo San Marcellino 10, 80138 Napoli, Italy}

\begin{document}

\title{\documenttitle}

\author{Roberta Calabrese}
\email{rcalabrese@na.infn.it}
\affiliation{\UNINA}
\affiliation{\INFN}
\author{Jacob Gunn}
\email{jacobwilliam.gunn@unina.it}
\affiliation{\UNINA}
\affiliation{\INFN}
\author{Gennaro Miele}
\email{gennaro.miele@unina.it}
\affiliation{\UNINA}
\affiliation{\INFN}
\affiliation{\SSM}
\author{Stefano Morisi}
\email{smorisi@na.infn.it}
\affiliation{\UNINA}
\affiliation{\INFN}
\author{Samiran Roy}
\email{samroy@na.infn.it}
\affiliation{\INFN}
\author{Pietro Santorelli}
\email{Pietro.Santorelli@na.infn.it}
\affiliation{\UNINA}
\affiliation{\INFN}

\begin{abstract}
Neutrino-nucleus coherent scattering measurements by the COHERENT collaboration provide us with a unique capability to test various beyond the standard model scenarios. In this work, we constrain scalar leptoquarks (LQs) using the COHERENT data. LQs arise in many extensions of the Standard Model (SM). Generally, the mass of the LQs is assumed to be very high to avoid the bounds from proton decay. However, there are low-scale LQ models which prohibit proton decay by construction. We consider two electroweak doublet scalar LQ models with hypercharge Y=1/6, and Y=7/6 and provide the bounds in the plane of the Yukawa coupling and the mass of LQ.  We also compare the bounds on LQs coming from various other experiments and find that the COHERENT one covers a wide range of LQ masses from MeV to TeV and in certain regions the constraints are competitive with the others.
\end{abstract}

\maketitle

\section{Introduction}

The detection of coherent elastic neutrino-nucleus scattering (CE$\nu$NS) was finally achieved by the COHERENT collaboration \cite{COHERENT:2017ipa} more than forty years after it was predicted by Freedman \cite{PhysRevD.9.1389}.
The cross-section of CE$\nu$NS events depends on the momentum transfer to the nucleus ($q^2$), and scales approximately as the square of the total number of protons (Z) and neutrons (N) in the target nucleus for low $q^2$. In the Standard Model (SM), CE$\nu$NS is mediated by the Z-boson, which couples preferentially to the neutron. Hence, the SM coherent cross section depends mainly on the square of N. The signals observed by the COHERENT collaboration show no significant deviation from the SM prediction, for CsI \cite{COHERENT:2017ipa,COHERENT:2021xmm} and Ar \cite{COHERENT:2020iec} nuclei.\\

The COHERENT data presents an opportunity to verify the SM parameters at low energy scales ($\sim $ 10 MeV); in this energy range, the weak mixing angles can be evaluated\,\cite{Miranda:2020tif,Cadeddu:2019eta,Cadeddu:2021ijh} as well as the root mean square radii of the neutron in Cesium, Iodide, and Argon\,\cite{Cadeddu:2017etk,Cadeddu:2019eta,Canas:2019fjw, Miranda:2020tif}. Furthermore, the COHERENT data is also sensitive to a plethora of new physics scenarios, such as extra mediators\,\cite{Dutta:2015vwa, Abdullah:2018ykz, Dutta:2019eml, Cadeddu:2020nbr, Flores:2020lji, Banerjee:2021laz, delaVega:2021wpx, AtzoriCorona:2022moj, Binh:2021iww, Bertuzzo:2021opb,Chakraborty:2021apc, DeRomeri:2022twg}, non-standard neutrino interactions (NSI) \cite{Miranda:2020tif,Barranco:2005yy,Scholberg:2005qs, Lindner:2016wff, Liao:2017uzy,Giunti:2019xpr,Denton:2020hop,Khan:2021wzy, Coloma:2022avw, Chatterjee:2022mmu}, generalized neutrino interactions\,\cite{Dutta:2015nlo,Flores:2021kzl}, neutrino magnetic moments\,\cite{Miranda:2019wdy}, non-unitarity of the leptonic mixing matrix\,\cite{Miranda:2020syh}, light sterile neutrinos\,\cite{Dutta:2015nlo, Kosmas:2017zbh, Miranda:2019skf, Miranda:2020syh}, dipole portal \cite{Dasgupta:2021fpn, Bolton:2021pey}, electromagnetic properties of neutrinos\,\cite{Cadeddu:2020lky}, fermionic dark matter\,\cite{Brdar:2018qqj, Dror:2019onn, Dror:2019dib}, light scalar coupling to both neutrino and quarks\,\cite{Farzan:2018gtr}, etc. For a wider review the reader is referred to\,\cite{DeRomeri:2022twg, Cadeddu:2020lky}. In this work, we focus on leptoquark (LQ) models.

LQs are particles beyond the SM, carrying both lepton and baryon numbers, that in general can be scalar or vector. Here we focus on the scalar LQs case which is less constrained compared to the vector one \cite{Dey:2015eaa, Valencia:1994cj, Smirnov:2007hv}. Indeed vector LQs couplings are fixed from the gauge structure and there is less freedom. They arise in a large variety of beyond the SM scenarios, such as grand unified theories \cite{PhysRevLett.32.438,FRITZSCH1975193, Preda:2022izo}, the Pati-Salam model \cite{PhysRevD.10.275}, extended technicolor models \cite{FARHI1981277,GABRIEL1994336, Andersen:2011yj}, etc. In general, the mass of such LQs are expected to be close to the GUT scale to avoid proton decay. There are however models with low scale LQs \cite{BUCHMULLER1987442,Belyaev:2005ew,Dorsner:2005fq,FileviezPerez:2013zmv} which forbid rapid proton decays. We consider scalar LQs which do not give rise to tree level interactions of the type ($q l q q$), so proton decay does not occur. 

The bounds on LQs via CEvNS using the effective field theory approach is discussed in \cite{Crivellin:2021bkd}. 
In contrast, we utilize the most recent data of CsI, and the data of Ar to derive the exclusion limits on scalar LQs via modification to the SM \cns event rate in the presence of LQs over a wide mass range from MeV to TeV scales. We also mention interesting features of the constraints and compare our results with existing bounds on LQs. Additional constraints on LQs can be found in \cite{Davidson:1993qk, Crivellin:2021egp}. However, we mention only those constraints which are relevant to our scenarios.

The structure of this paper is as follows: section \ref{Model} details our model and notation and provides the relevant part of the Lagrangian that mediates the coherent process. In section \ref{signal_coherent}, we describe CE$\nu$NS SM and LQ signal predictions. Section \ref{analysis} contains the $\chi^2$ analysis. In section \ref{Results}, we report on our exclusion limits and some discussion. Finally, section \ref{conclusions} contains our conclusions. 

\section{Models}
\label{Model}

\begin{table}[tbh!]
\begin{tabular}{c|c|c|}
\cline{2-3}
                                                     & $SU(3)_c\times SU(2)_L \times U(1)_Y$ & LQ                                        \\ \hline\hline
\multicolumn{1}{|c|}{$(\overline{Q^c}_L)^iL^j_L$}    & ${\bf (3, 3, -1/3)}$                  & $S_3{\bf (\overline{3}, 3, 1/3)}$         \\ \hline
\multicolumn{1}{|c|}{$(\overline{Q^c}_L)^iL^j_L$}    & ${\bf (3, 1, -1/3)}$                  & $S_1{\bf (\overline{3}, 1, 1/3)}$         \\ \hline
\multicolumn{1}{|c|}{$(\overline{u^c}_R)^i\ell_R^j$} & ${\bf (3, 1, -1/3)}$                  & $S_1{\bf (\overline{3}, 1, 1/3)}$         \\ \hline
\multicolumn{1}{|c|}{$\overline{u}_R^iL^j_L$}        & ${\bf (\overline{3}, 2, -7/6)}$       & $R_2{\bf (3, 2, 7/6)}$         \\ \hline
\multicolumn{1}{|c|}{$\overline{\ell}_R^iQ_L^j$}     & ${\bf (\overline{3}, 2, -7/6)}$       & $R_2{\bf (3, 2, 7/6)}$                    \\ \hline
\multicolumn{1}{|c|}{$\overline{d}_R^iL^j_L$}        & ${\bf (\overline{3}, 2, -1/6)}$       & $\tilde{R}_2{\bf (3, 2, 1/6)}$            \\ \hline
\multicolumn{1}{|c|}{$(\overline{d^c}_R)^i\ell_R^j$} & ${\bf (3, 1, -4/3)}$                  & $\tilde{S}_1{\bf (\overline{3}, 1, 4/3)}$ \\ \hline\hline
\end{tabular}
\caption{\label{tab:LQ} The first column corresponds to the list of $q\ell$ interaction terms. The second column shows the quantum numbers of the $q\ell$ terms. In the third column, we report the LQ quantum numbers. For more details, see Ref.\,\cite{Dorsner:2016wpm}}.
\end{table}

LQs couple to quarks and leptons. Therefore, LQ quantum numbers are not arbitrary and must be fixed accordingly. In Tab\,\ref{tab:LQ}, we give the $SU(2)_L \times U(1)_Y$ quantum numbers of all possible terms involving quarks, leptons and a scalar LQ. The above table does not exhaustively report the possible LQ interactions which also include ``diquark" type interactions for $S_3$, $S_1$, and $\tilde{S}_1$. $\tilde{R}_2$ and $ R_2$ do not contain any ``diquark" type interactions which cause proton decay \cite{Dorsner:2016wpm}. Note that the term $H \tilde R \tilde R \tilde R $, may cause proton decay($p \rightarrow \pi^+ + \pi^- + e^- +\nu \nu$)\cite{Arnold:2012sd}, however in Refs.\,\cite{Crivellin:2021ejk, Murgui:2021bdy, Dorsner:2022twk} it is shown that such an operator vanishes for $\Tilde{R}$. For simplicity, we consider each LQ separately and do not analyze the scenarios with two LQs simultaneously. The SM Lagrangian is minimally extended to include the Yukawa interaction of the LQs with the quark and lepton fields. The relevant part of the Lagrangian that contributes in the neutrino-nucleon coherent process is given by
\begin{equation}
    \begin{split}
        \mathcal{L}^{\Delta_1} &\supset -y_{ij}^{(1)} \bar{d}^i_{R} \Tilde{R}_2 L^j_{L} + h.c. \\
        &= -y_{ij}^{(1)} \bar{d}^i P_L \ell^j \Delta^{2/3}_1 -y_{ij}^{(1)} \bar{d}^i P_L \nu^j \Delta^{-1/3}_1 + h.c.
    \end{split}
    \label{L1}
\end{equation}
and 
\begin{equation}
    \begin{split}
       \mathcal{L}^{\Delta_2} &\supset -y_{ij}^{(2)} \bar{u}^i_{R} R_2 L^j_{L} + h.c. \\
       &= -y_{ij}^{(2)} \bar{u}^i P_L \ell^j \Delta^{5/3}_2 -y_{ij}^{(2)} \bar{u}^i P_L \nu^j \Delta^{2/3}_2 + h.c.\,,
    \end{split}
    \label{L2}
\end{equation}
where $\Tilde{R}_2 = (\Delta^{2/3}_1 ,\Delta^{-1/3}_1 )^T$ and $R_2 = (\Delta^{5/3}_2 ,\Delta^{2/3}_2 )^T$ under $SU(2)_L$, $i,j$ are generation indices, $d^i$ and $u^i$ are the down-type and up-type quark fields, $l^j_L$ are the lepton doublets $L^j_L = \begin{pmatrix}\nu^j & \ell^j \end{pmatrix} ^T$, $P_{L,R}$,  are the usual left and right chiral projection operators and $y'$s are the Yukawa coupling matrices. The masses of LQs in the doublet structure could be different in general. However, substantial mass splitting generates large corrections to $T$ parameter \cite{Davidson:2010uu}. To evade such constraint, the small mass splitting ($<<m_{W}$) is preferable. In that scenario, the constraint on $\Delta^{2/3}_1 \, (\Delta^{5/3}_2)$ is directly bound the parameters of the Yukawa matrix $y^1 \,(y^2)$ as well. Since CE$\nu$NS involves quark mass eigenstates and neutrino interaction eigenstates, the coupling matrices we are interested in are $\tilde{y}^{(1)}= V_d^\dagger \, y^{(1)}$ and $\tilde{y}^{(2)}= V_u^\dagger \, y^{(2)}$ where $V_d$ ($V_u$) are arbitrary, unitary matrices which rotate the quark fields into their mass basis. Therefore, since $y^{(1)}$ ($y^{(2)}$) are arbitrary matrices, $\tilde{ y}^{(1)}$ ($\tilde{ y}^{(2)}$) are also completely arbitrary \textit{a priori}. To avoid having Flavour Changing Neutral Current (FCNC) problems at tree level, we consider only couplings with the first generation of quarks. This choice does not affect our prediction for the CE$\nu$NS since it only involves the valence quarks, hence only the first generation. 

In principle, one should also include terms like 
\begin{equation}
\mathcal{L} \supset y_{ij}^{(3)}\overline{\ell}^i_R (R_2^*)^a (Q_L^j)^a  + h.c.\,.
\label{eq:L-piondecay}
\end{equation}
Such terms do not contribute to CE$\nu$NS but can modify the pion decay. There are strong constraint on LQ coming from pion decay if we consider simultaneously both the Eq.\,\ref{L2} and Eq.\ref{eq:L-piondecay} type operators \cite{Davidson:1993qk}. We assume the coupling in Eq.\ref{eq:L-piondecay} to be very small and explore the COHERENT data to constrain the couplings in Eqns.\,\ref{L1} and \,\ref{L2}. 
\section{Signal Prediction in COHERENT}
\label{signal_coherent}
In this section, we present the signal prediction at COHERENT for both the SM and LQs scenarios. 

At the COHERENT experiment, the neutrino fluxes come from the Spallation Neutron Source (SNS). Prompt pion decay ($\pi^+ \rightarrow \mu^+ +\nu_{\mu}$) at rest  and the subsequently delayed decay of muons ($\mu^+ \rightarrow e^+ + \nu_e + \bar{\nu}_{\mu}$) produce three different neutrino fluxes. The $\nu_{\mu}$ reach the detector within $\sim 1.5 ~ \mu$s  after protons-on-target while the $\bar{\nu}_{\mu}$ and $\nu_e$ fluxes arrive in a comparatively longer time interval ($\sim 10~\mu$s). The differential fluxes for neutrinos are as follows :
\begin{eqnarray}
\dfrac{dN_{\nu_{\mu}}}{dE} = \eta \, \delta \, \Big(E- \dfrac{m^2_{\pi}-m^2_{\mu}}{2m_{\pi}}\Big)
\end{eqnarray}  
\begin{eqnarray}
\dfrac{dN_{\bar{\nu}_{\mu}}}{dE} = \eta \, \dfrac{64E^2}{m^3_{\mu}} \, \Big( \dfrac{3}{4} - \dfrac{E}{m_{\mu}}\Big )
\end{eqnarray} 
\begin{eqnarray}
\dfrac{dN_{\nu_{e}}}{dE} = \eta \, \dfrac{192E^2}{m^3_{\mu}} \, \Big( \dfrac{1}{2} - \dfrac{E}{m_{\mu}}\Big ),
\end{eqnarray}
where $m_{\mu}$ and $m_{\pi}$ are the muon and pion masses respectively, $\eta=r\, \rm{N_{POT}}/4\pi L^2$ is a normalization factor, $\rm{N_{POT}}$ is the total number of protons on target (POT), $r$ corresponds to neutrinos per flavour produced for each POT, and L measures the distance between the source and the detector. For the CsI detector $r$ = 0.08, $\rm{N_{POT}} = 3.198 \times 10^{23}$ and L = 19.3 m\,\cite{COHERENT:2021xmm}\, while for Ar $ r = (9 \pm 0.9)\times 10^{-2}$, $\rm{N_{POT}} = 13.7\times 10^{22}$, L =27.5 m. \\

 The {\it SM contribution} to the  \cns cross-section for a neutrino ($\nu_l$) as a function of the nuclear recoil kinetic energy ($T_{nr}$) is given by
\begin{eqnarray}\label{SM}
\dfrac{d \sigma_{\nu_l - N}}{dT_{nr}} (E,T_{nr})= \dfrac{G^2_F M}{\pi} \Big (1-\dfrac{MT_{nr}}{2E^2}\Big) Q^2_{l,SM}.
\label{eq:sigma_SM}
\end{eqnarray}
In the above, $l$ is the neutrino flavour index, $M$ is the mass of the detector nucleus with $Z$ protons and $N$ neutrons, and $E$ is the energy of incoming neutrino, $G_F$ is the Fermi constant,  and 
\begin{equation}
Q^2_{l,SM}=[g^p_V(\nu_l)ZF_Z(|\vec{q}|^2) + g^n_V N F_N (|\vec{q}|^2)]^2. 
\label{QSM} 
\end{equation}
Considering the radiative corrections in the minimal subtraction, $\rm{\overline{MS}}$ scheme, the values of the $g^p_V$ and $g^n_V$ couplings are\,\cite{Erler:2013xha, Cadeddu:2020lky}
\begin{equation}
g^p_V(\nu_e)=0.0401 ,\,\, g^p_V(\nu_{\mu})=0.0318,\,\, g^n_V=-0.5094.
\label{g_p_g_n}
\end{equation}
The proton and neutron distributions in the nucleus are represented by the form factors ($F_Z(|\vec{q}|^2)$ and $F_N(|\vec{q}|^2)$) which depend on the three-momentum transfer $|\vec{q}|\simeq \sqrt{2MT_{nr}}$. We use the Helm parameterization \cite{PhysRev.104.1466}  for the form factors\footnote{ The form factor of various materials relevant to CE$\nu$NS events using the large-scale nuclear shell model is discussed in \cite{Hoferichter:2020osn} .}
\begin{eqnarray}
F(|\vec{q}|^2)=3\dfrac{j_1 (|\vec{q}|R_0)}{|\vec{q}|R_0} e^{-|\vec{q}|^2s^2/2},
\end{eqnarray}
where $j_1(x)=\sin(x)/x^2- \cos(x)/x$ is the spherical Bessel function and $s=0.9$ fm, and the rms radius $R$ is related to $R_0$ by $R^2_{p, n} = 3R^2_0/5 + 3s^2$. Helm's form factors are effectively identical to the Fermi parameterization for our purposes \cite{Giunti:2019xpr}. We take $R_p(\rm{Cs})=4.804 \,\rm{fm} $ and $R_p(\rm{I})=4.749 \, \rm{fm}$
as the rms radius of the proton\,\cite{Fricke:1995zz, Angeli:2013epw} in CsI. For Argon, $R_p(\rm{Ar})=3.448 \,\rm{fm}$. The neutron rms radius is not known with good accuracy for CsI or Ar. In our analysis, we take the values $R_n(\rm{Cs})=5.01 $ fm and $R_n(\rm{I})= 4.94$ fm \cite{Bender:1999yt} For Argon, we use the predicted difference between the neutron and proton radii, the so-called neutron skin. Since most models predict a skin around $0.1$fm \cite{Reinhard:1995zz,Niksic:2008vp,Payne:2019wvy} we take $R_n(\rm{Ar})=3.55 \,\rm{fm}$. If the transfer of momentum is such that $|\vec{q}|R_p \gtrsim 1$ and $|\vec{q}|R_n \gtrsim 1$, the scattering loses coherence.\\
\begin{figure}[tbh!]\label{scatter}
    \centering
    \includegraphics[width = 0.20 \textwidth]{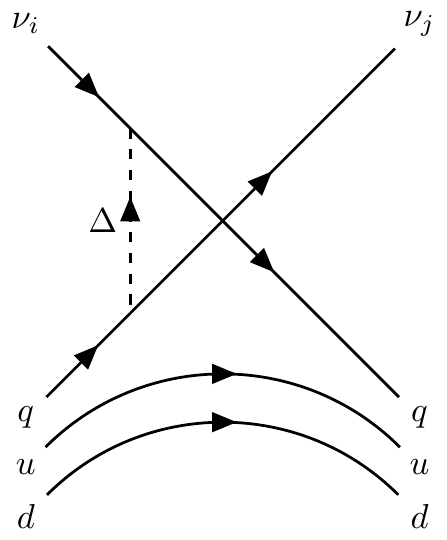}
    \includegraphics[width = 0.20 \textwidth]{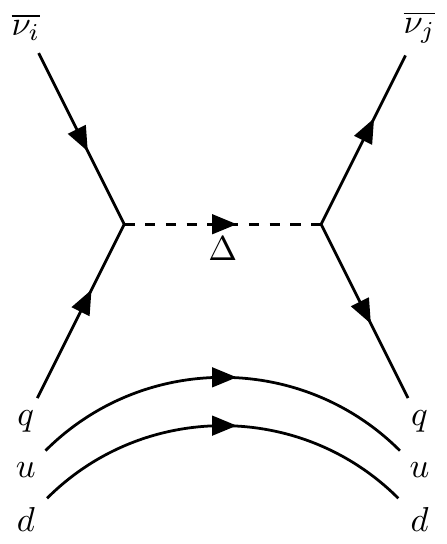}
   \caption{\em \label{fig:Coherent} Tree level processes involving the Leptoquark which modifies the SM prediction for the CE$\nu$NS cross-section. The final state neutral lepton flavor is not experimentally distinguishable.}
\end{figure}
The number of events in the i-th bin of nuclear-recoil energy is given by
\begin{eqnarray}
\label{eq:en_spectra}\nonumber
 N_i^{\mathrm{CE}\nu\mathrm{NS}} = N(\mathcal{N}) \int^{T_{\mathrm{nr}}^{i+1}}_{T_{\mathrm{nr}}^{i}} dT_{\mathrm{nr}}\,A(T_{\mathrm{nr}}) \int^{E_{\mathrm{max}}}_{E_{\mathrm{min}}} dE \\
 \sum_{\nu=\nu_e,\nu_\mu,\overline{\nu}_\mu} \frac{dN_\nu}{dE} \dfrac{d\sigma_{\nu\text{-}\mathcal{N}}}{d T_\mathrm{nr}}(E,T_\mathrm{nr})\,,
\end{eqnarray}
where $E_{\mathrm{min}} = \sqrt{MT_\mathrm{nr}/2}$, $E_{\mathrm{max}} = m_\mu/2 \sim 52.8$ MeV, and $A(T_{\mathrm{nr}})$ is the energy-dependent reconstruction efficiency. The nuclear recoil energy $T_{\mathrm{nr}}\,[\mathrm{keV}]$ is converted into the corresponding  electron recoil energy $T_{ee}\,[\mathrm{keV}]$. For CsI 
\begin{equation}
\label{eq:qf}
T_{ee} =  (1-\alpha_3) f_Q (T_{\mathrm{nr}})T_{\mathrm{nr}}\,,
\end{equation}
where $f_Q$ is the quenching factor, which we take from \cite{Collar:2019ihs}, and $\alpha_3$ is a nuisance parameter quantifying the uncertainty on $f_Q$. For Argon
\begin{equation}
T_{ee} = f_Q (T_{\mathrm{nr}})T_{\mathrm{nr}}\,.
\end{equation}
where the quenching factor is parameterised as
\begin{equation}
f_Q = (0.246 \pm 0.006\rm{keV}) + ((7.8 \pm 0.9) \times 10^{-4})T_{nr}\,,
\end{equation}
following \cite{COHERENT:2020iec}, for $0\,\rm{keV} < T_{nr} < 125\,\rm{keV}$. For larger energies $f_Q$ is a constant. $N(\mathcal{N})$ is the total number of atoms present in the active detector mass ($M_{\rm det}$) and $N(\mathcal{N}) = N_\mathrm{A}\,M_{\mathrm{det}}/M_{\mathcal{N}}$, where $N_\mathrm{A}$ is Avogadro's number and $M_{\mathcal{N}}$ is the molar mass of the detector material. For CsI we take  $M_{\mathrm{det}}=14.6$ kg and $M_{\mathrm{CsI}}=259.8$ g/mol and for Ar we take $M_{\mathrm{det}}=24$ kg and $M_{\mathrm{Ar}}=39.96$ for Ar$^{40}$.
\vspace{3mm}

In the presence of LQs the standard \cns cross section will be modified. Processes contributing to the CE$\nu$NS are shown Fig.\,\ref{fig:Coherent}. In this figure we can see that we have vertices with a neutrino, LQ, and a quark. As the typical momentum transfer to the nucleus is the order of a few tens of MeV, these diagrams can be represented by the effective four fermion interaction for $m_{\Delta} \gtrapprox 10$ MeV like
\begin{eqnarray}
\mathcal{L}^{\Delta}_{\rm{eff}} = \dfrac{y^2}{m^2_{\Delta}}\, (\bar{\psi}_N P_L \nu) \, (\bar{\nu}P_R \psi_N)\,.
\label{Eff.eq}
\end{eqnarray}
where $\psi_N$ is either a $u$ or $d$ quark field. However, we need to factorize this diagram into a neutrino current and a hadronic current to compute the CE$\nu$NS events. This can be achieved through Fierz transformations giving the following effective interaction 

\begin{eqnarray}
\mathcal{L}^{\Delta}_{\rm{eff}} \sim -\dfrac{y^2}{2 m^2_{\Delta}}\, (\bar{\psi}_N \gamma^{\mu}P_R \psi_N) \, (\bar{\nu}\gamma_{\mu}P_L  \nu) .
\label{Eff.eq1}
\end{eqnarray} 
%This is defined in detail in Appendix\,\ref{app}. 

Writing the effective interaction in this way, with a $V-A$ axial structure, allows us to write down the \cns cross section including the contribution from LQs
\begin{equation}
    \frac{d \sigma_{\nu_i - N}}{dT_{nr}} (E,T_{nr})= \dfrac{G^2_F M}{\pi} \left (1-\dfrac{MT_{nr}}{2E^2}\right) Q^2_{i,\, k}
    \label{eq:sigma_LQ}
\end{equation}
\begin{equation}
Q^2_{i,\, k} = \left(\left(Q_{i,\,SM}+ Q_{ii,\, \Delta_k}\right)^2+\sum_{j\neq i} Q^2_{ij,\, \Delta_k}\right)
\label{eq:Q_ik}
\end{equation}
where 
\begin{equation}
\begin{split}
    &Q_{ij,\, \Delta_1} = \frac{\tilde{y}_{1i}^{(1)}\tilde{y}_{1j}^{(1)}}{4\sqrt{2}\,G_F}\, \frac{Z F_Z(|\vec{q}|^2)+ 2\,N F_N(|\vec{q}|^2)}{|\vec{q}|^2+ m_{\Delta_1^{-1/3}}^2} ,\\
    &Q_{ij,\, \Delta_2} = \frac{\tilde{y}_{1i}^{(2)}\tilde{y}_{1j}^{(2)}}{4\sqrt{2}\,G_F}\, \frac{2\,Z F_Z(|\vec{q}|^2)+ N F_N(|\vec{q}|^2)}{|\vec{q}|^2+ m_{\Delta_2^{2/3}}^2} .
\end{split}
\label{eq:Qdelta}
\end{equation}
\begin{figure}[bh!]
    \centering
    \includegraphics[width = 0.5 \textwidth]{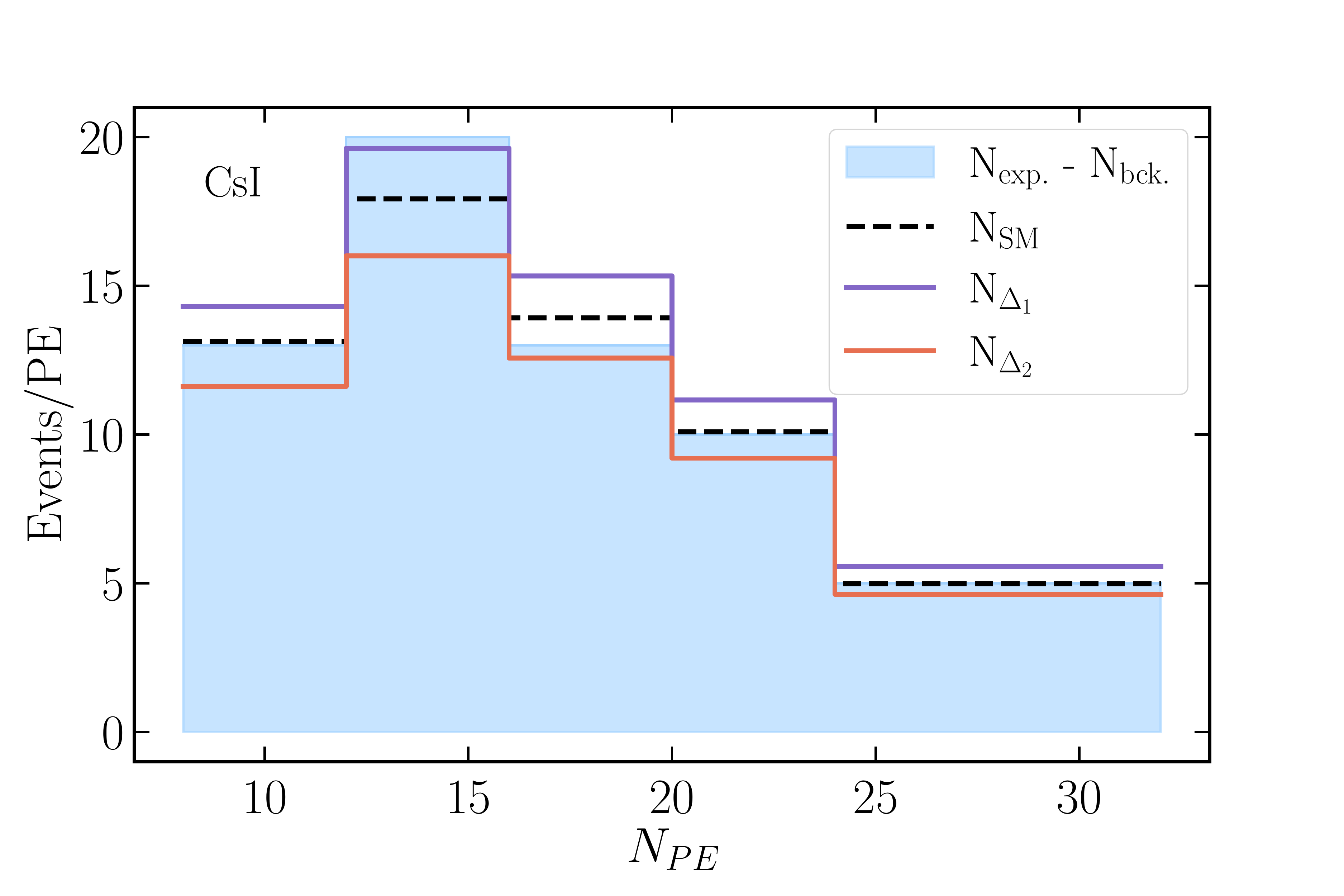}
   \caption{\em \label{fig:Events} The plot shows the number of events vs the photo-electron number. The difference between experimental and the background  event rate is presented in cyan. The dashed black line represents the SM event prediction, the violet solid line corresponds to the prediction including $\Delta_1^{-1/3}$, the orange solid line is the prediction for $\Delta_2^{2/3}$. The theoretical predictions for the case A (see Tab.\,\ref{tab:scenarios}) are obtained assuming $m_{\Delta} = 1$GeV and $g_\Delta = 3.5\cdot 10^{-3}$.}
\end{figure}
Note that typical $U'(1)$ based models\,\cite{Cadeddu:2020nbr}  consider the processes like $N+\nu_i\to N+\nu_i$ for CE$\nu$NS. On the contrary here we have $N+\nu_i\to N+\nu_j$ where $i $ can be different from $ j$ as can be seen from Fig.\,\ref{fig:Coherent}. 

\begin{table*}[th!]
\begin{tabular}{|c||c|c|c|}
\hline
\hline
\backslashbox{LQ}{Case} & A & B & C \\
\hline
\hline
% FIRST ROW 
$\Delta_1^{-1/3}$ 
    & $\tilde{y}^{(1)}= \left(\begin{array}{ccc} g_{\Delta_1} & g_{\Delta_1} & 0 \\ 0 & 0& 0\\ 0 & 0& 0\end{array}\right)$ 
    & $\tilde{y}^{(1)}= \left(\begin{array}{ccc} 0 & g_{\Delta_1} & 0 \\ 0 & 0& 0\\ 0 & 0& 0\end{array}\right)$ 
    & $\tilde{y}^{(1)}= \left(\begin{array}{ccc} g_{\Delta_1} & 0 & 0 \\ 0 & 0& 0\\ 0 & 0& 0\end{array}\right)$ \\
\hline
% SECOND ROW 
$\Delta_2^{2/3}$ 
    & $\tilde{y}^{(2)}= \left(\begin{array}{ccc} g_{\Delta_2} & g_{\Delta_2} & 0 \\ 0 & 0& 0\\ 0 & 0& 0\end{array}\right)$ 
    & $\tilde{y}^{(2)}= \left(\begin{array}{ccc} 0 & g_{\Delta_2} & 0 \\ 0 & 0& 0\\ 0 & 0& 0\end{array}\right)$ 
    & $\tilde{y}^{(2)}=\left(\begin{array}{ccc} g_{\Delta_2} &0 & 0 \\ 0 & 0& 0\\ 0 & 0& 0\end{array}\right)$\\
\hline
\hline
\end{tabular}
\caption{\label{tab:scenarios} In the table we report the benchmark cases studied in this paper.}
\end{table*}
%=========%
% RESULTS %
%=========%
\section{Analysis}\label{analysis}
\subsection{CsI}

Using Eqns.\,(\ref{eq:sigma_SM}, \ref{eq:sigma_LQ}, \ref{eq:Q_ik}, \ref{eq:Qdelta}), we compute the theoretical prediction for the event number of CE$\nu$NS assuming only the SM or assuming the existence of either $\Delta_1$ or $\Delta_2$. As mentioned above, we are considering the LQs that couple only to the first  generation of quarks since CE$\nu$NS is sensitive only to valence quarks. Moreover, this prevents us from having FCNC at tree level mediated by LQs. For each LQ, we consider three possible scenarios. $\Delta_k$ couples to both $\nu_e$ and $\nu_\mu$ (case A) or $\Delta_k$ couples only to one of $\nu_\mu$ and $\nu_e$ (Case B and C) as summarized in Tab.\,\ref{tab:scenarios}. It is possible to consider other possibilities, but they are less conservative. 

To compare our predictions with the measured event rate, we need to write our prediction as a function of the photo-electron number instead of the nuclear recoil energy. In Ref.\,\cite{COHERENT:2018imc}, it is shown that 
\begin{equation}
N_{PE} = Y_T T_{ee}\, ,
\end{equation}
where $Y_T = 13.35 \,N_{PE}/keV$ is the light yield of the phototubes. Eq.\,\ref{eq:en_spectra} can be rewritten in terms of $N_{PE}$ as 
\begin{eqnarray}
\label{eq:en_spectra_PE}\nonumber
 N_i^{\mathrm{CE}\nu\mathrm{NS}} = N(\mathcal{N}) \int^{N_{PE}^{i+1}}_{N_{PE}^{i}} dN_{PE}\,f(N_{PE}) \int^{E_{\mathrm{max}}}_{E_{\mathrm{min}}} dE \\
 \sum_{\nu=\nu_e,\nu_\mu,\overline{\nu}_\mu} \frac{dN_\nu}{dE} \dfrac{d\sigma_{\nu\text{-}\mathcal{N}}}{d T_\mathrm{nr}}(E,T_\mathrm{nr})\,,
\end{eqnarray}

where $f(N_{PE})$ is the detector efficiency in terms of the photo-electron content defined as\,\cite{COHERENT:2021xmm}
\begin{equation}
f(x) = \frac{a}{1+\exp(-k (x-x_0))} + d\, ,
\end{equation}
where 
\begin{equation}
\begin{split}
&a = 1.32045\pm 0.02345 \, ,\\
&k = 0.2859792\pm 0.000613\, , \\
&x_0 =10.8646\pm 1.0186 \, ,\\
&d = -0.333322\pm 0.023042\,.
\end{split}
\end{equation}

In Fig.\,\ref{fig:Events}, we compare the COHERENT \cns event rate, in cyan, to the SM prediction in black, and our case A model predictions in violet and orange. The plot shows the event per photo-electron. To get the number of events measured, one has to multiply the height and width of each bin. Since $Q_{i, SM}$ and $Q_{ii, \Delta_k}$ have opposite sign, the resulting event rate can be higher or smaller than the SM one depending on $(g_{\Delta_{1,2}}, m_{\Delta_{1,2}})$

To constrain the models, we perform a $\chi^2$ analysis analogous to that in Ref.\,\cite{Cadeddu:2020nbr}
\begin{equation}
\begin{split}
\chi^2 = \min_{\boldsymbol{\alpha}}\sum_{i = 4}^{15}&\left[\left(\frac{N_i^{\rm exp} - (1+\alpha_1)N_i^{CE\nu NS}-(1+\alpha_2)B_i}{\sigma_i}\right)^2\right. \\
        & \left. + \sum_j \left(\frac{\alpha_j}{\sigma_{\alpha_j}}\right)^2\right]\,.
\end{split}
\label{eq:chi_csi}
\end{equation}
For each bin $i$, we indicate with $N_i^{\rm exp}$ the measured event number reported in Ref.\,\cite{COHERENT:2021xmm}, $N_i^{CE\nu NS}$ is the expected event rate, $B_i$ is the background event rate\,\cite{COHERENT:2021xmm}, and $\sigma_i$ is the statistical uncertainty taken to be equal to the square root of the event number. As done in Refs.\,\cite{Cadeddu:2019eta, Cadeddu:2020nbr}, we consider only the energy bins from 8 to 32 photo-electrons because they span the recoil kinetic energy range covered by the most recent quenching factor measurement performed at Chicago-3\,\cite{Collar:2019ihs}. We are considering 3 nuisance parameters $\boldsymbol{\alpha} = (\alpha_1, \alpha_2, \alpha_3)$ which quantify the systematic uncertainty of the signal rate, background rate, and quenching factor respectively. The corresponding standard deviations are $\sigma_{\boldsymbol{\alpha}} = (0.112, 0.25, 0.051)$\,\cite{COHERENT:2017ipa}. 

The choices we made for the Yukawa couplings prevent us from having constraints from FCNC and B-physics. Moreover, we can overcome the constraints from LFV considering a mass splitting between the members of the doublets.

\subsection{Argon}
%\begin{figure}[th!]
%    \centering
%    \includegraphics[width = 0.49 \textwidth]{Plots/Event_Ar.pdf}
%   \caption{\em \label{fig:Event_argon} In this plot we show the number of events vs the electron recoil %energy. The experimental number of events minus the background is reported in cyan. We report the SM %prediction as a black dashed line. Assuming Case A (see Tab.\,\ref{tab:scenarios}), we report the event %number obtained for $\Delta_2^{2/3}$ in purple and the event number obtained for $\Delta_1^{-1/3}$ in %orange. Both prediction were obtained assuming $m_{\Delta} = 1$GeV and $g_\Delta = 4\cdot 10^{-3}$. }
%\end{figure}

Analogously to the preceding analysis for CsI, using Eq.\,(\ref{eq:en_spectra}) we can obtain the theoretical prediction for the CE$\nu$NS event rate in the Argon detector assuming only the SM or assuming the existence of either $\Delta_1$ or $\Delta_2$. The energy-dependent reconstruction coefficient for Argon is also taken from analysis A in Ref.\,\cite{COHERENT:2020iec}.

We perform the following $\chi^2$ analysis
\begin{equation}
\begin{split}
\chi^2_{Ar} =& \sum_{i=1}^{12} \left( \frac{ N^{exp}_i - (1+\beta_{1}) N^{CE\nu NS}_i- B_i}{\sigma_i}\right)^2 \\
&+\sum_j \left(\frac{\beta_{j}}{\sigma_{\beta_{j}}}\right)^2 
\end{split}
\label{eq:chi_square_ar}
\end{equation}
where $B_i = (1+\beta_{2})B^{PBRN}_i + (1+\beta_{3}) B^{LBRN}_i$. PBRN means Prompt Beam-Related Background and LBRN is the Late-Beam-Related Neutron Background, so that $B_i^{PBRN},B_i^{LBRN}$ are the estimated number of PBRN and LBRN events in the i$^{\rm th}$ energy bin. In the above\,\cite{COHERENT:2020iec}
\begin{equation}
\sigma^2_i = (\sigma^{exp}_i)^2 + (\sigma_{BRNES}(B_i^{PBRN} + B_i^{LBRN}))^2 
\end{equation}
where $\sigma_{BRNES} = 1.7\%$ and $\sigma_{\boldsymbol{\beta}} = (0.132, 0.32, 1)$\,\cite{COHERENT:2020iec}. In this way, the uncertainty on the Beam Related Neutron Energy Shape (BRNES) is distributed over the energy bins in an uncorrelated manner, and $\beta_1,\,\beta_2,\,\beta_3$ are the nuisance parameters which quantify the uncertainty on the signal rate and PBRN and LBRN background event rates.

\section{Results}\label{Results}
\begin{figure*}[tbh!]
    \centering
    \includegraphics[width = 0.49 \textwidth]{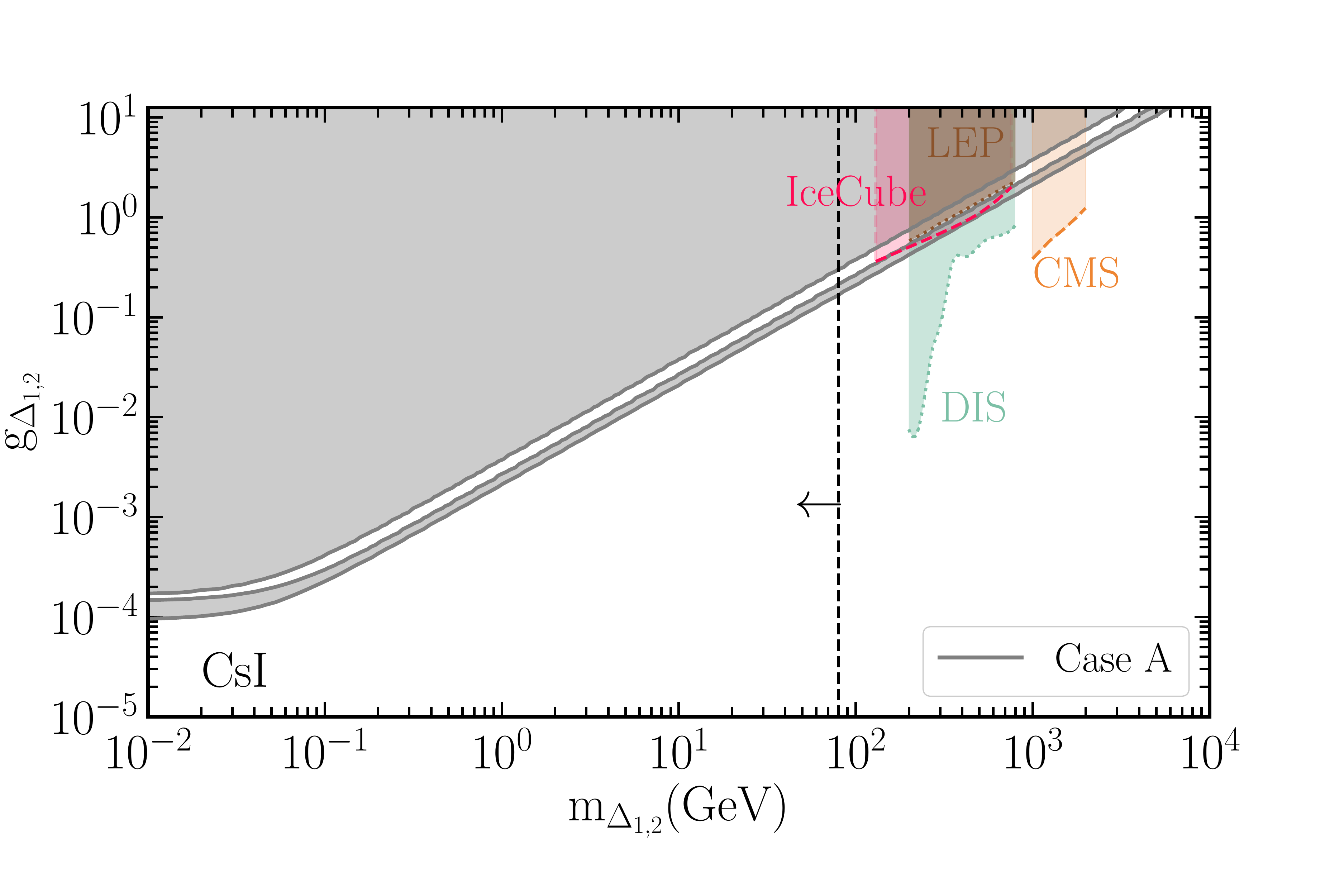}
    \includegraphics[width = 0.49\textwidth]{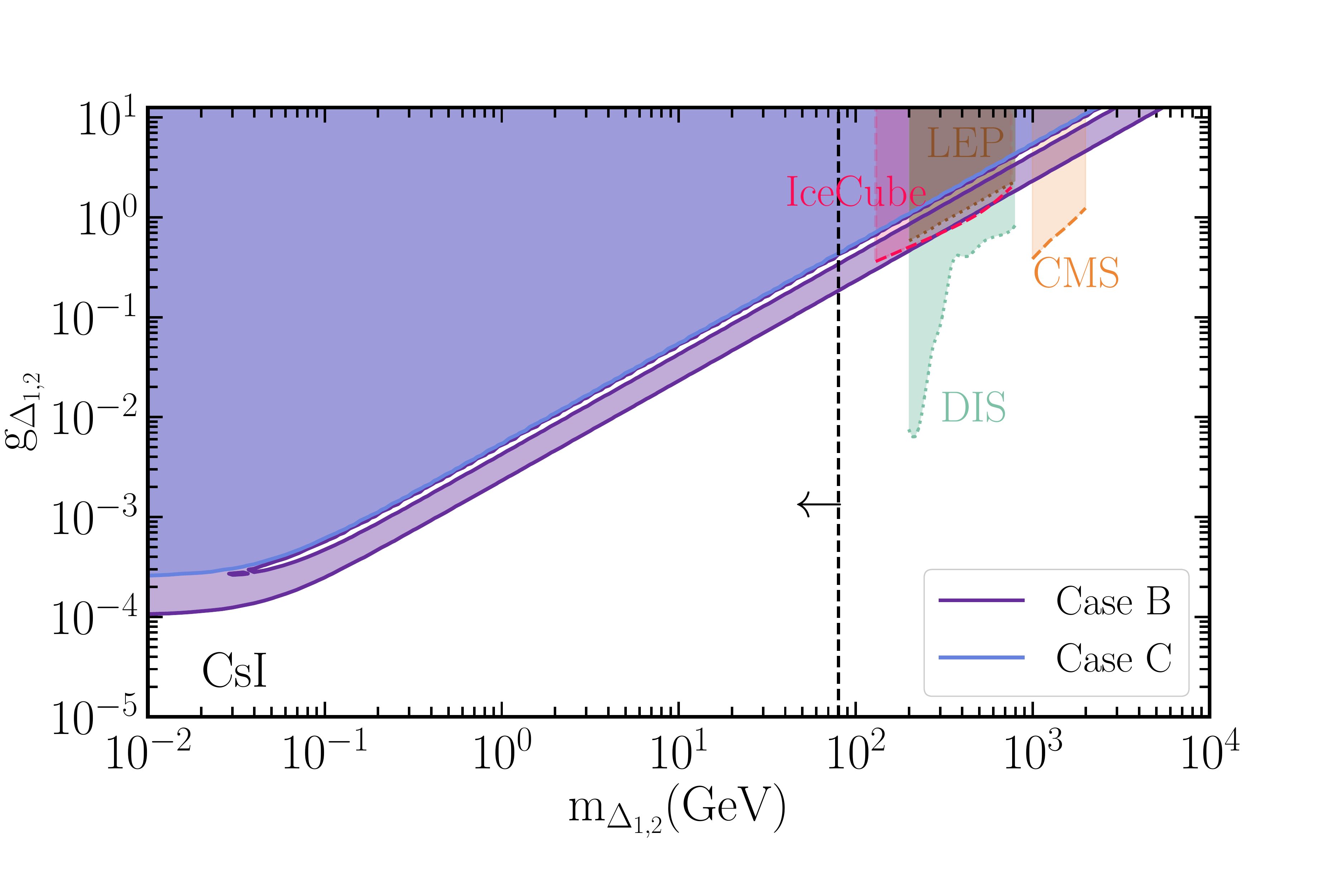}
     \includegraphics[width = 0.49 \textwidth]{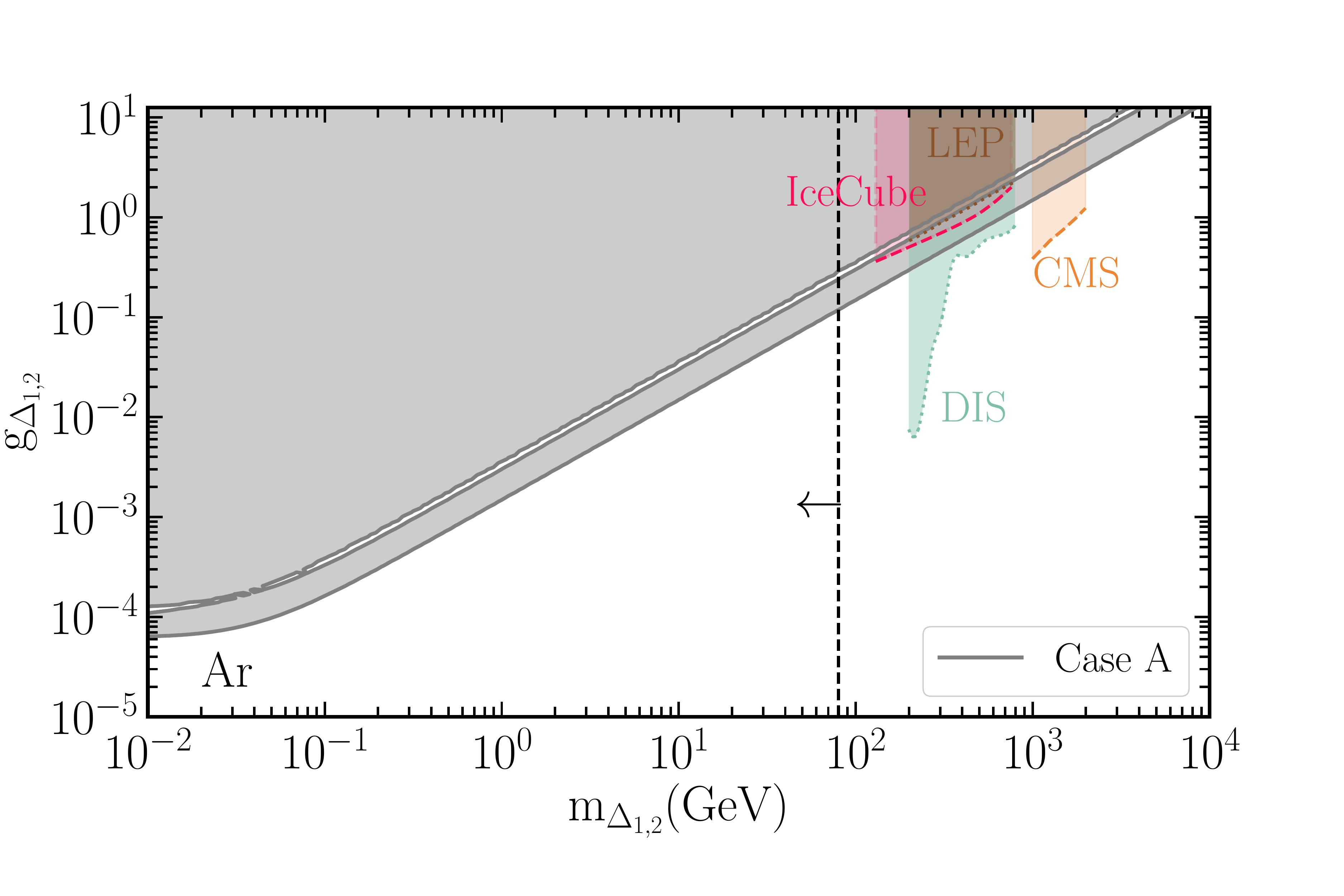}
    \includegraphics[width = 0.49\textwidth]{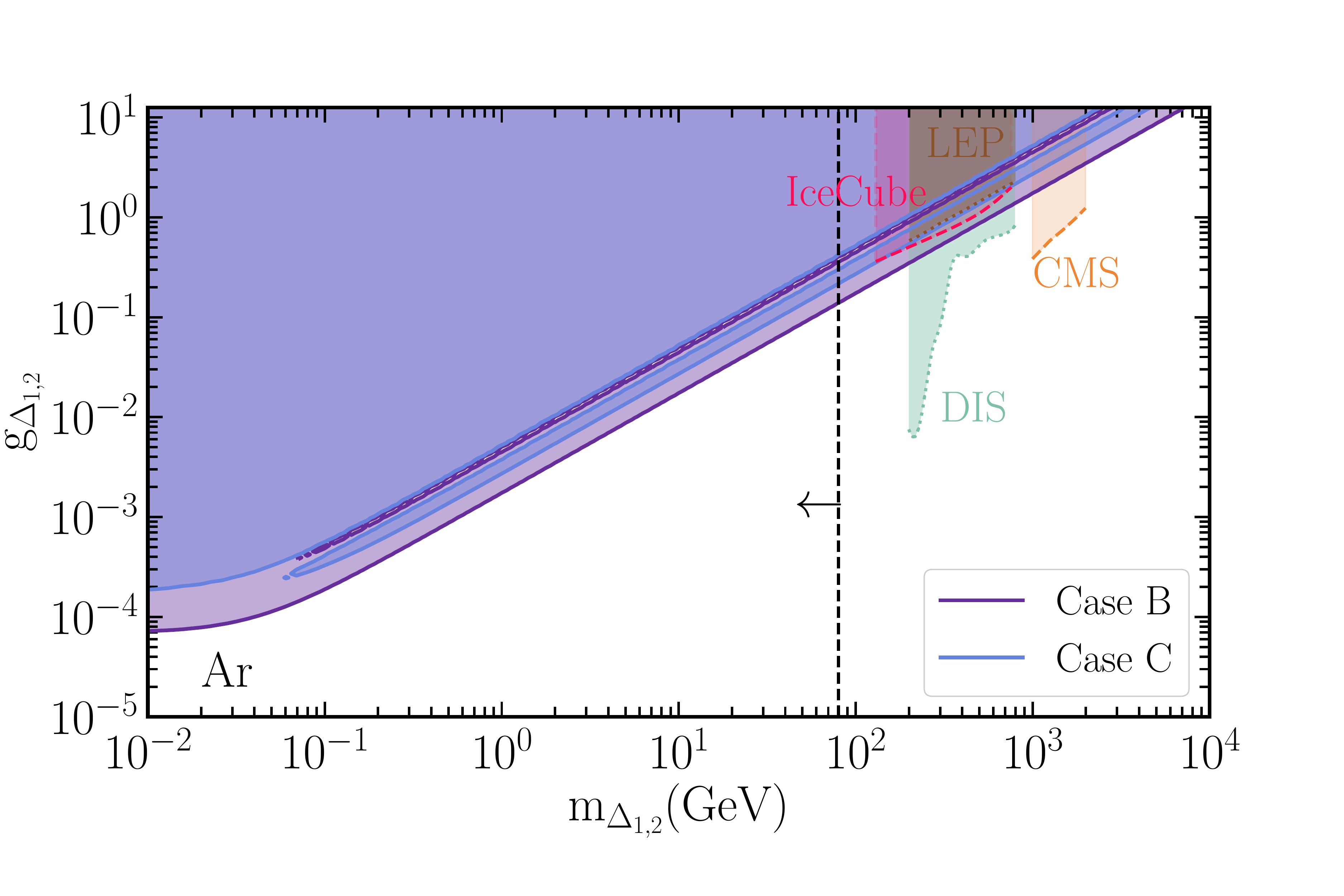}
   \caption{\em \label{fig:PPlots} Constraints obtained in the plane $g_\Delta-m_\Delta$. In the left panel, the constraints on Case A are shown, while on the right we present the constraints on Case B and C. In the upper half we show the constraints obtained using CsI, and in the lower half those from Argon. In solid colored lines we have reported the constraints at 90\% C.L. obtained using our analysis. The dashed colored lines show the constraints for $\Delta_1^{-1/3}$ and $\Delta_2^{2/3}$ from CMS (in orange) \cite{CMS:2021far} and IceCube (in pink) \cite{Dey:2015eaa}. The dotted brown and green lines represent the constraints for $\Delta_1^{2/3}$ and $\Delta_2^{5/3}$. Such constraints are taken from Ref.\,\cite{Taxil:1999pf}.}
\end{figure*}
We show the constraints on the Yukawa coupling strength as a function of the LQ masses in Fig.\,\ref{fig:PPlots} for the cases described in Tab. \ref{tab:scenarios}. The top and bottom panels correspond to CsI and Ar respectively, whereas the left and right panels depict the constraint on case A only, and cases B and C respectively. The constraints on $\Delta_{1,2}$ differ only at the percent level, therefore we report them as a single solid line. Expectedly, the most stringent constraints apply to scenario A since all of the $\nu_{\mu}\, , \bar{\nu}_{\mu} $ and $\nu_e$ fluxes contribute to the \cns events rate, while only the $\nu_{\mu,\Bar{\mu}}$ and only the $\nu_e$ participate in case B and C respectively. For very low LQs masses (around 10 MeV), the bounds become insensitive to $m_{\Delta}$ as in this region, the cross section depends mainly on the momentum transfer ($q^2$). In each plot, we show some existing constraints on LQs for comparison. The dashed lines are used for the bounds which constrain $\Delta_1^{-1/3}$ or $\Delta_2^{2/3}$, while the dotted lines correspond to those which are obtained for the other member of the doublets.  The orange dashed line corresponds to constraints from CMS taken from Ref.\,\cite{CMS:2021far}, for which the LQs ($\Delta$) are pair produced via gluon fusion and the constraints are obtained assuming $BR(\Delta\to q+\nu) = 1$. A similar bound is reported using the pair production of asymmetric LQs in Ref.\,\cite{Dorsner:2022ibm}. IceCube also provide constraints on LQs which we show in pink \,\cite{Dey:2015eaa}, and are discussed in Refs.\,\cite{Dey:2017ede, Becirevic:2018uab}. The brown and green dotted lines show the LEP and Deep Inelastic Scattering bounds taken from Ref.\,\cite{Taxil:1999pf}, which constrain the charged lepton element of the LQ doublet. A large mass splitting of the doublet components leads to sizable correction to $T$ parameter\,\cite{Davidson:2010uu}. Hence, in the limit of degenerate LQ doublet masses, those bounds would apply to our scenarios. The low mass ($\lesssim 80$ GeV) LQs are severely constrained by the LEP searches via the unsuppressed channel $e^+e^-\to \gamma^* \to \Delta^+\Delta^-$. Note that in Ref.\,\cite{ALEPH:2013htx}, the constraint is obtained for charged Higgs but this is also applicable to charged LQs. The bound is indicated by the black dashed line.
Constraints may also arise from other experimental observations, like $K^0-\overline{K}^0$ or $D^0-\overline{D}^0$ oscillation or in general B physics, but such constraints do not apply due to the Yukawa structure considered. 

\begin{figure}[tbh]
    \centering
    \includegraphics[width = 0.49 \textwidth]{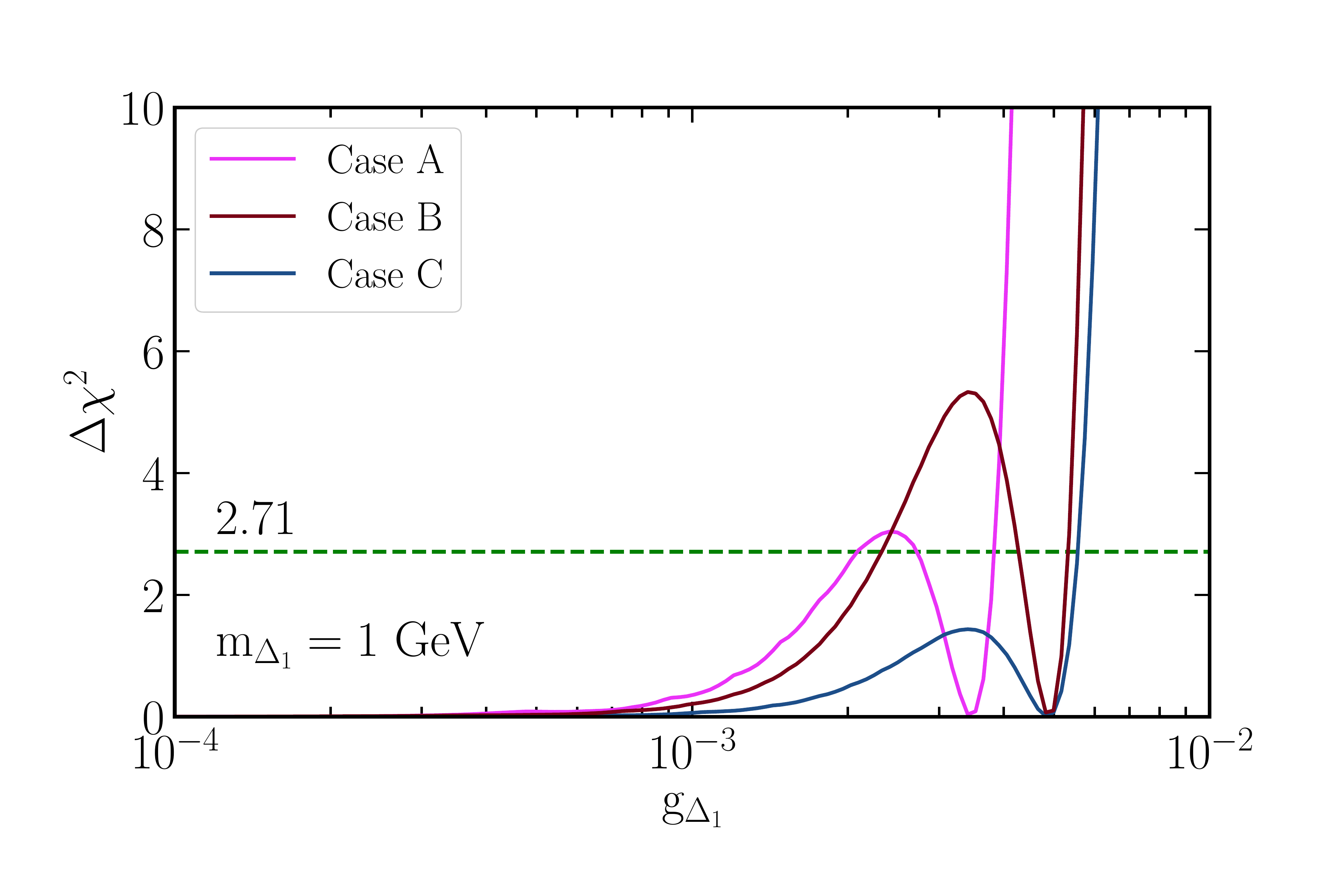}
   \caption{\em \label{fig:chi} $\Delta \chi^2$ vs $g_{\Delta_1}$ for $m_{\Delta_1} = 1$ GeV for CsI. In magenta, red and blue we report Case A, B and C respectively (see Tab.\,\ref{tab:scenarios}). The green line refers to $\Delta  \chi^2 =2.71$.}
\end{figure}

The constrained regions are not continuous due to a degeneracy of the SM and LQ contributions. In Fig.\,\ref{fig:chi} we show the behaviour of $\Delta \chi^2$ for $\Delta_1^{-1/3}$ obtained using Eq.\,\ref{eq:chi_csi} to illustrate the shape of the constraints. The magenta, red, and blue lines corresponds to cases A, B, and C respectively for $m_{\Delta_1} = 1$ GeV. By looking at Eqs.\,\ref{QSM}, \ref{g_p_g_n} and taking the values of the form factors to one, we obtain the value of $Q_{i\,SM}\simeq -N/2$. On the other hand, $Q_{ij, \Delta_K}>0$. Therefore, there is a degenerate point where the total new charge approaches $N/2$. From Eq. \ref{eq:Q_ik}, we get  
\begin{equation}
    \begin{split}
        &{\rm for\, case\,A}:\left(-\frac{N}{2}+Q_{ii, \Delta_K}\right)^2 +\sum_{i\neq j} Q_{ij, \Delta_K}^2 \simeq \left(\frac{N}{2}\right)^2,\\
        &{\rm for \,case\,B\,and\,C}:\left(-\frac{N}{2}+Q_{ii, \Delta_K}\right)^2\simeq \left(\frac{N}{2}\right)^2\,,
    \end{split}
     \label{eq: chi_sm}
\end{equation}
which almost reproduces the SM prediction. As the ratio $g_\Delta/m_\Delta$ increases from zero, the predicted event rate including LQs is initially decreasing from the SM prediction. At a certain point the prediction equals the SM one (see Eq.\,\ref{eq: chi_sm}), and the $\Delta \chi^2 = 0$. This is the degeneracy which causes the discontinuity in our constraints. 
%If it is bigger (like in Case A and B, see Fig.\,\ref{fig:chi}), then the obtained bounds are characterized by two regions. If it is smaller (like in Case C), we get only get one point for $g_\Delta$ for each $m_\Delta$. 

\section{Conclusions}\label{conclusions}
In this work, we utilize the neutrino-nucleus coherent elastic scattering data to put bounds on scalar LQs over a wide mass range from MeV to TeV scales. 
We consider scalar LQs with hypercharge Y=1/6 and Y=7/6. These LQs do not have any ``diquark'' couplings which contribute to the proton decay. Hence effectively the LQ masses could be well below the GUT scale. We use a Fierz transformation to recast the effective four-fermion interaction vertex into neutrino-neutrino and quark-quark currents. This allows us to calculate the hadronic matrix elements for the coherent process. Only the valence quarks (u and d) of the nuclei participate in neutrino-nucleus coherent scattering. Therefore, we can constrain the specific structure of the Yukawa matrix that couples to the first generation of quarks, and neutrinos. Our choices of couplings evade bounds from FCNCs and LFV. Hence, COHERENT measurements allow us to probe this particular Yukawa structure. We also show the constraints coming from LHC, IceCube, LEP, and DIS. These constraints are comparable with our bounds for some mass regions. Placing constraints on LQs over a wide mass range is the unique capability of coherent neutrino-nucleus scattering measurements.

\begin{center}
\textbf{Acknowledgments}
\end{center} 
We thank  Carlo Giunti, Ujjal Kumar Dey, and Marco Chianese for the useful discussion. 

This work was partially supported by the research grant number 2017W4HA7S ``NAT-NET: Neutrino and Astroparticle Theory Network'' under the program PRIN 2017 funded by the Italian Ministero dell'Universit\`a e della Ricerca (MUR).

\bibliography{Bibliography}

\begin{thebibliography}{83}%
\makeatletter
\providecommand \@ifxundefined [1]{%
 \@ifx{#1\undefined}
}%
\providecommand \@ifnum [1]{%
 \ifnum #1\expandafter \@firstoftwo
 \else \expandafter \@secondoftwo
 \fi
}%
\providecommand \@ifx [1]{%
 \ifx #1\expandafter \@firstoftwo
 \else \expandafter \@secondoftwo
 \fi
}%
\providecommand \natexlab [1]{#1}%
\providecommand \enquote  [1]{``#1''}%
\providecommand \bibnamefont  [1]{#1}%
\providecommand \bibfnamefont [1]{#1}%
\providecommand \citenamefont [1]{#1}%
\providecommand \href@noop [0]{\@secondoftwo}%
\providecommand \href [0]{\begingroup \@sanitize@url \@href}%
\providecommand \@href[1]{\@@startlink{#1}\@@href}%
\providecommand \@@href[1]{\endgroup#1\@@endlink}%
\providecommand \@sanitize@url [0]{\catcode `\\12\catcode `\$12\catcode
  `\&12\catcode `\#12\catcode `\^12\catcode `\_12\catcode `\%12\relax}%
\providecommand \@@startlink[1]{}%
\providecommand \@@endlink[0]{}%
\providecommand \url  [0]{\begingroup\@sanitize@url \@url }%
\providecommand \@url [1]{\endgroup\@href {#1}{\urlprefix }}%
\providecommand \urlprefix  [0]{URL }%
\providecommand \Eprint [0]{\href }%
\providecommand \doibase [0]{http://dx.doi.org/}%
\providecommand \selectlanguage [0]{\@gobble}%
\providecommand \bibinfo  [0]{\@secondoftwo}%
\providecommand \bibfield  [0]{\@secondoftwo}%
\providecommand \translation [1]{[#1]}%
\providecommand \BibitemOpen [0]{}%
\providecommand \bibitemStop [0]{}%
\providecommand \bibitemNoStop [0]{.\EOS\space}%
\providecommand \EOS [0]{\spacefactor3000\relax}%
\providecommand \BibitemShut  [1]{\csname bibitem#1\endcsname}%
\let\auto@bib@innerbib\@empty
%</preamble>
\bibitem [{\citenamefont {Akimov}\ \emph {et~al.}(2017)\citenamefont {Akimov}
  \emph {et~al.}}]{COHERENT:2017ipa}%
  \BibitemOpen
  \bibfield  {author} {\bibinfo {author} {\bibfnamefont {D.}~\bibnamefont
  {Akimov}} \emph {et~al.} (\bibinfo {collaboration} {COHERENT}),\ }\href
  {\doibase 10.1126/science.aao0990} {\bibfield  {journal} {\bibinfo  {journal}
  {Science}\ }\textbf {\bibinfo {volume} {357}},\ \bibinfo {pages} {1123}
  (\bibinfo {year} {2017})},\ \Eprint {http://arxiv.org/abs/1708.01294}
  {arXiv:1708.01294 [nucl-ex]} \BibitemShut {NoStop}%
\bibitem [{\citenamefont {Freedman}(1974)}]{PhysRevD.9.1389}%
  \BibitemOpen
  \bibfield  {author} {\bibinfo {author} {\bibfnamefont {D.~Z.}\ \bibnamefont
  {Freedman}},\ }\href {\doibase 10.1103/PhysRevD.9.1389} {\bibfield  {journal}
  {\bibinfo  {journal} {Phys. Rev. D}\ }\textbf {\bibinfo {volume} {9}},\
  \bibinfo {pages} {1389} (\bibinfo {year} {1974})}\BibitemShut {NoStop}%
\bibitem [{\citenamefont {Akimov}\ \emph {et~al.}(2022)\citenamefont {Akimov}
  \emph {et~al.}}]{COHERENT:2021xmm}%
  \BibitemOpen
  \bibfield  {author} {\bibinfo {author} {\bibfnamefont {D.}~\bibnamefont
  {Akimov}} \emph {et~al.} (\bibinfo {collaboration} {COHERENT}),\ }\href
  {\doibase 10.1103/PhysRevLett.129.081801} {\bibfield  {journal} {\bibinfo
  {journal} {Phys. Rev. Lett.}\ }\textbf {\bibinfo {volume} {129}},\ \bibinfo
  {pages} {081801} (\bibinfo {year} {2022})},\ \Eprint
  {http://arxiv.org/abs/2110.07730} {arXiv:2110.07730 [hep-ex]} \BibitemShut
  {NoStop}%
\bibitem [{\citenamefont {Akimov}\ \emph {et~al.}(2021)\citenamefont {Akimov}
  \emph {et~al.}}]{COHERENT:2020iec}%
  \BibitemOpen
  \bibfield  {author} {\bibinfo {author} {\bibfnamefont {D.}~\bibnamefont
  {Akimov}} \emph {et~al.} (\bibinfo {collaboration} {COHERENT}),\ }\href
  {\doibase 10.1103/PhysRevLett.126.012002} {\bibfield  {journal} {\bibinfo
  {journal} {Phys. Rev. Lett.}\ }\textbf {\bibinfo {volume} {126}},\ \bibinfo
  {pages} {012002} (\bibinfo {year} {2021})},\ \Eprint
  {http://arxiv.org/abs/2003.10630} {arXiv:2003.10630 [nucl-ex]} \BibitemShut
  {NoStop}%
\bibitem [{\citenamefont {Miranda}\ \emph
  {et~al.}(2020{\natexlab{a}})\citenamefont {Miranda}, \citenamefont
  {Papoulias}, \citenamefont {Sanchez~Garcia}, \citenamefont {Sanders},
  \citenamefont {T\'ortola},\ and\ \citenamefont {Valle}}]{Miranda:2020tif}%
  \BibitemOpen
  \bibfield  {author} {\bibinfo {author} {\bibfnamefont {O.~G.}\ \bibnamefont
  {Miranda}}, \bibinfo {author} {\bibfnamefont {D.~K.}\ \bibnamefont
  {Papoulias}}, \bibinfo {author} {\bibfnamefont {G.}~\bibnamefont
  {Sanchez~Garcia}}, \bibinfo {author} {\bibfnamefont {O.}~\bibnamefont
  {Sanders}}, \bibinfo {author} {\bibfnamefont {M.}~\bibnamefont {T\'ortola}},
  \ and\ \bibinfo {author} {\bibfnamefont {J.~W.~F.}\ \bibnamefont {Valle}},\
  }\href {\doibase 10.1007/JHEP05(2020)130} {\bibfield  {journal} {\bibinfo
  {journal} {JHEP}\ }\textbf {\bibinfo {volume} {05}},\ \bibinfo {pages} {130}
  (\bibinfo {year} {2020}{\natexlab{a}})},\ \bibinfo {note} {[Erratum: JHEP 01,
  067 (2021)]},\ \Eprint {http://arxiv.org/abs/2003.12050} {arXiv:2003.12050
  [hep-ph]} \BibitemShut {NoStop}%
\bibitem [{\citenamefont {Cadeddu}\ \emph
  {et~al.}(2020{\natexlab{a}})\citenamefont {Cadeddu}, \citenamefont {Dordei},
  \citenamefont {Giunti}, \citenamefont {Li},\ and\ \citenamefont
  {Zhang}}]{Cadeddu:2019eta}%
  \BibitemOpen
  \bibfield  {author} {\bibinfo {author} {\bibfnamefont {M.}~\bibnamefont
  {Cadeddu}}, \bibinfo {author} {\bibfnamefont {F.}~\bibnamefont {Dordei}},
  \bibinfo {author} {\bibfnamefont {C.}~\bibnamefont {Giunti}}, \bibinfo
  {author} {\bibfnamefont {Y.~F.}\ \bibnamefont {Li}}, \ and\ \bibinfo {author}
  {\bibfnamefont {Y.~Y.}\ \bibnamefont {Zhang}},\ }\href {\doibase
  10.1103/PhysRevD.101.033004} {\bibfield  {journal} {\bibinfo  {journal}
  {Phys. Rev. D}\ }\textbf {\bibinfo {volume} {101}},\ \bibinfo {pages}
  {033004} (\bibinfo {year} {2020}{\natexlab{a}})},\ \Eprint
  {http://arxiv.org/abs/1908.06045} {arXiv:1908.06045 [hep-ph]} \BibitemShut
  {NoStop}%
\bibitem [{\citenamefont {Cadeddu}\ \emph
  {et~al.}(2021{\natexlab{a}})\citenamefont {Cadeddu}, \citenamefont
  {Cargioli}, \citenamefont {Dordei}, \citenamefont {Giunti}, \citenamefont
  {Li}, \citenamefont {Picciau}, \citenamefont {Ternes},\ and\ \citenamefont
  {Zhang}}]{Cadeddu:2021ijh}%
  \BibitemOpen
  \bibfield  {author} {\bibinfo {author} {\bibfnamefont {M.}~\bibnamefont
  {Cadeddu}}, \bibinfo {author} {\bibfnamefont {N.}~\bibnamefont {Cargioli}},
  \bibinfo {author} {\bibfnamefont {F.}~\bibnamefont {Dordei}}, \bibinfo
  {author} {\bibfnamefont {C.}~\bibnamefont {Giunti}}, \bibinfo {author}
  {\bibfnamefont {Y.~F.}\ \bibnamefont {Li}}, \bibinfo {author} {\bibfnamefont
  {E.}~\bibnamefont {Picciau}}, \bibinfo {author} {\bibfnamefont {C.~A.}\
  \bibnamefont {Ternes}}, \ and\ \bibinfo {author} {\bibfnamefont {Y.~Y.}\
  \bibnamefont {Zhang}},\ }\href {\doibase 10.1103/PhysRevC.104.065502}
  {\bibfield  {journal} {\bibinfo  {journal} {Phys. Rev. C}\ }\textbf {\bibinfo
  {volume} {104}},\ \bibinfo {pages} {065502} (\bibinfo {year}
  {2021}{\natexlab{a}})},\ \Eprint {http://arxiv.org/abs/2102.06153}
  {arXiv:2102.06153 [hep-ph]} \BibitemShut {NoStop}%
\bibitem [{\citenamefont {Cadeddu}\ \emph {et~al.}(2018)\citenamefont
  {Cadeddu}, \citenamefont {Giunti}, \citenamefont {Li},\ and\ \citenamefont
  {Zhang}}]{Cadeddu:2017etk}%
  \BibitemOpen
  \bibfield  {author} {\bibinfo {author} {\bibfnamefont {M.}~\bibnamefont
  {Cadeddu}}, \bibinfo {author} {\bibfnamefont {C.}~\bibnamefont {Giunti}},
  \bibinfo {author} {\bibfnamefont {Y.~F.}\ \bibnamefont {Li}}, \ and\ \bibinfo
  {author} {\bibfnamefont {Y.~Y.}\ \bibnamefont {Zhang}},\ }\href {\doibase
  10.1103/PhysRevLett.120.072501} {\bibfield  {journal} {\bibinfo  {journal}
  {Phys. Rev. Lett.}\ }\textbf {\bibinfo {volume} {120}},\ \bibinfo {pages}
  {072501} (\bibinfo {year} {2018})},\ \Eprint
  {http://arxiv.org/abs/1710.02730} {arXiv:1710.02730 [hep-ph]} \BibitemShut
  {NoStop}%
\bibitem [{\citenamefont {Canas}\ \emph {et~al.}(2020)\citenamefont {Canas},
  \citenamefont {Garces}, \citenamefont {Miranda}, \citenamefont {Parada},\
  and\ \citenamefont {Sanchez~Garcia}}]{Canas:2019fjw}%
  \BibitemOpen
  \bibfield  {author} {\bibinfo {author} {\bibfnamefont {B.~C.}\ \bibnamefont
  {Canas}}, \bibinfo {author} {\bibfnamefont {E.~A.}\ \bibnamefont {Garces}},
  \bibinfo {author} {\bibfnamefont {O.~G.}\ \bibnamefont {Miranda}}, \bibinfo
  {author} {\bibfnamefont {A.}~\bibnamefont {Parada}}, \ and\ \bibinfo {author}
  {\bibfnamefont {G.}~\bibnamefont {Sanchez~Garcia}},\ }\href {\doibase
  10.1103/PhysRevD.101.035012} {\bibfield  {journal} {\bibinfo  {journal}
  {Phys. Rev. D}\ }\textbf {\bibinfo {volume} {101}},\ \bibinfo {pages}
  {035012} (\bibinfo {year} {2020})},\ \Eprint
  {http://arxiv.org/abs/1911.09831} {arXiv:1911.09831 [hep-ph]} \BibitemShut
  {NoStop}%
\bibitem [{\citenamefont {Dutta}\ \emph
  {et~al.}(2016{\natexlab{a}})\citenamefont {Dutta}, \citenamefont {Mahapatra},
  \citenamefont {Strigari},\ and\ \citenamefont {Walker}}]{Dutta:2015vwa}%
  \BibitemOpen
  \bibfield  {author} {\bibinfo {author} {\bibfnamefont {B.}~\bibnamefont
  {Dutta}}, \bibinfo {author} {\bibfnamefont {R.}~\bibnamefont {Mahapatra}},
  \bibinfo {author} {\bibfnamefont {L.~E.}\ \bibnamefont {Strigari}}, \ and\
  \bibinfo {author} {\bibfnamefont {J.~W.}\ \bibnamefont {Walker}},\ }\href
  {\doibase 10.1103/PhysRevD.93.013015} {\bibfield  {journal} {\bibinfo
  {journal} {Phys. Rev. D}\ }\textbf {\bibinfo {volume} {93}},\ \bibinfo
  {pages} {013015} (\bibinfo {year} {2016}{\natexlab{a}})},\ \Eprint
  {http://arxiv.org/abs/1508.07981} {arXiv:1508.07981 [hep-ph]} \BibitemShut
  {NoStop}%
\bibitem [{\citenamefont {Abdullah}\ \emph {et~al.}(2018)\citenamefont
  {Abdullah}, \citenamefont {Dent}, \citenamefont {Dutta}, \citenamefont
  {Kane}, \citenamefont {Liao},\ and\ \citenamefont
  {Strigari}}]{Abdullah:2018ykz}%
  \BibitemOpen
  \bibfield  {author} {\bibinfo {author} {\bibfnamefont {M.}~\bibnamefont
  {Abdullah}}, \bibinfo {author} {\bibfnamefont {J.~B.}\ \bibnamefont {Dent}},
  \bibinfo {author} {\bibfnamefont {B.}~\bibnamefont {Dutta}}, \bibinfo
  {author} {\bibfnamefont {G.~L.}\ \bibnamefont {Kane}}, \bibinfo {author}
  {\bibfnamefont {S.}~\bibnamefont {Liao}}, \ and\ \bibinfo {author}
  {\bibfnamefont {L.~E.}\ \bibnamefont {Strigari}},\ }\href {\doibase
  10.1103/PhysRevD.98.015005} {\bibfield  {journal} {\bibinfo  {journal} {Phys.
  Rev. D}\ }\textbf {\bibinfo {volume} {98}},\ \bibinfo {pages} {015005}
  (\bibinfo {year} {2018})},\ \Eprint {http://arxiv.org/abs/1803.01224}
  {arXiv:1803.01224 [hep-ph]} \BibitemShut {NoStop}%
\bibitem [{\citenamefont {Dutta}\ \emph {et~al.}(2019)\citenamefont {Dutta},
  \citenamefont {Liao}, \citenamefont {Sinha},\ and\ \citenamefont
  {Strigari}}]{Dutta:2019eml}%
  \BibitemOpen
  \bibfield  {author} {\bibinfo {author} {\bibfnamefont {B.}~\bibnamefont
  {Dutta}}, \bibinfo {author} {\bibfnamefont {S.}~\bibnamefont {Liao}},
  \bibinfo {author} {\bibfnamefont {S.}~\bibnamefont {Sinha}}, \ and\ \bibinfo
  {author} {\bibfnamefont {L.~E.}\ \bibnamefont {Strigari}},\ }\href {\doibase
  10.1103/PhysRevLett.123.061801} {\bibfield  {journal} {\bibinfo  {journal}
  {Phys. Rev. Lett.}\ }\textbf {\bibinfo {volume} {123}},\ \bibinfo {pages}
  {061801} (\bibinfo {year} {2019})},\ \Eprint
  {http://arxiv.org/abs/1903.10666} {arXiv:1903.10666 [hep-ph]} \BibitemShut
  {NoStop}%
\bibitem [{\citenamefont {Cadeddu}\ \emph
  {et~al.}(2021{\natexlab{b}})\citenamefont {Cadeddu}, \citenamefont
  {Cargioli}, \citenamefont {Dordei}, \citenamefont {Giunti}, \citenamefont
  {Li}, \citenamefont {Picciau},\ and\ \citenamefont
  {Zhang}}]{Cadeddu:2020nbr}%
  \BibitemOpen
  \bibfield  {author} {\bibinfo {author} {\bibfnamefont {M.}~\bibnamefont
  {Cadeddu}}, \bibinfo {author} {\bibfnamefont {N.}~\bibnamefont {Cargioli}},
  \bibinfo {author} {\bibfnamefont {F.}~\bibnamefont {Dordei}}, \bibinfo
  {author} {\bibfnamefont {C.}~\bibnamefont {Giunti}}, \bibinfo {author}
  {\bibfnamefont {Y.~F.}\ \bibnamefont {Li}}, \bibinfo {author} {\bibfnamefont
  {E.}~\bibnamefont {Picciau}}, \ and\ \bibinfo {author} {\bibfnamefont
  {Y.~Y.}\ \bibnamefont {Zhang}},\ }\href {\doibase 10.1007/JHEP01(2021)116}
  {\bibfield  {journal} {\bibinfo  {journal} {JHEP}\ }\textbf {\bibinfo
  {volume} {01}},\ \bibinfo {pages} {116} (\bibinfo {year}
  {2021}{\natexlab{b}})},\ \Eprint {http://arxiv.org/abs/2008.05022}
  {arXiv:2008.05022 [hep-ph]} \BibitemShut {NoStop}%
\bibitem [{\citenamefont {Flores}\ \emph {et~al.}(2020)\citenamefont {Flores},
  \citenamefont {Nath},\ and\ \citenamefont {Peinado}}]{Flores:2020lji}%
  \BibitemOpen
  \bibfield  {author} {\bibinfo {author} {\bibfnamefont {L.~J.}\ \bibnamefont
  {Flores}}, \bibinfo {author} {\bibfnamefont {N.}~\bibnamefont {Nath}}, \ and\
  \bibinfo {author} {\bibfnamefont {E.}~\bibnamefont {Peinado}},\ }\href
  {\doibase 10.1007/JHEP06(2020)045} {\bibfield  {journal} {\bibinfo  {journal}
  {JHEP}\ }\textbf {\bibinfo {volume} {06}},\ \bibinfo {pages} {045} (\bibinfo
  {year} {2020})},\ \Eprint {http://arxiv.org/abs/2002.12342} {arXiv:2002.12342
  [hep-ph]} \BibitemShut {NoStop}%
\bibitem [{\citenamefont {Banerjee}\ \emph {et~al.}(2021)\citenamefont
  {Banerjee}, \citenamefont {Dutta},\ and\ \citenamefont
  {Roy}}]{Banerjee:2021laz}%
  \BibitemOpen
  \bibfield  {author} {\bibinfo {author} {\bibfnamefont {H.}~\bibnamefont
  {Banerjee}}, \bibinfo {author} {\bibfnamefont {B.}~\bibnamefont {Dutta}}, \
  and\ \bibinfo {author} {\bibfnamefont {S.}~\bibnamefont {Roy}},\ }\href
  {\doibase 10.1103/PhysRevD.104.015015} {\bibfield  {journal} {\bibinfo
  {journal} {Phys. Rev. D}\ }\textbf {\bibinfo {volume} {104}},\ \bibinfo
  {pages} {015015} (\bibinfo {year} {2021})},\ \Eprint
  {http://arxiv.org/abs/2103.10196} {arXiv:2103.10196 [hep-ph]} \BibitemShut
  {NoStop}%
\bibitem [{\citenamefont {de~la Vega}\ \emph {et~al.}(2021)\citenamefont {de~la
  Vega}, \citenamefont {Flores}, \citenamefont {Nath},\ and\ \citenamefont
  {Peinado}}]{delaVega:2021wpx}%
  \BibitemOpen
  \bibfield  {author} {\bibinfo {author} {\bibfnamefont {L.~M.~G.}\
  \bibnamefont {de~la Vega}}, \bibinfo {author} {\bibfnamefont {L.~J.}\
  \bibnamefont {Flores}}, \bibinfo {author} {\bibfnamefont {N.}~\bibnamefont
  {Nath}}, \ and\ \bibinfo {author} {\bibfnamefont {E.}~\bibnamefont
  {Peinado}},\ }\href {\doibase 10.1007/JHEP09(2021)146} {\bibfield  {journal}
  {\bibinfo  {journal} {JHEP}\ }\textbf {\bibinfo {volume} {09}},\ \bibinfo
  {pages} {146} (\bibinfo {year} {2021})},\ \Eprint
  {http://arxiv.org/abs/2107.04037} {arXiv:2107.04037 [hep-ph]} \BibitemShut
  {NoStop}%
\bibitem [{\citenamefont {Atzori~Corona}\ \emph {et~al.}(2022)\citenamefont
  {Atzori~Corona}, \citenamefont {Cadeddu}, \citenamefont {Cargioli},
  \citenamefont {Dordei}, \citenamefont {Giunti}, \citenamefont {Li},
  \citenamefont {Picciau}, \citenamefont {Ternes},\ and\ \citenamefont
  {Zhang}}]{AtzoriCorona:2022moj}%
  \BibitemOpen
  \bibfield  {author} {\bibinfo {author} {\bibfnamefont {M.}~\bibnamefont
  {Atzori~Corona}}, \bibinfo {author} {\bibfnamefont {M.}~\bibnamefont
  {Cadeddu}}, \bibinfo {author} {\bibfnamefont {N.}~\bibnamefont {Cargioli}},
  \bibinfo {author} {\bibfnamefont {F.}~\bibnamefont {Dordei}}, \bibinfo
  {author} {\bibfnamefont {C.}~\bibnamefont {Giunti}}, \bibinfo {author}
  {\bibfnamefont {Y.~F.}\ \bibnamefont {Li}}, \bibinfo {author} {\bibfnamefont
  {E.}~\bibnamefont {Picciau}}, \bibinfo {author} {\bibfnamefont {C.~A.}\
  \bibnamefont {Ternes}}, \ and\ \bibinfo {author} {\bibfnamefont {Y.~Y.}\
  \bibnamefont {Zhang}},\ }\href {\doibase 10.1007/JHEP05(2022)109} {\bibfield
  {journal} {\bibinfo  {journal} {JHEP}\ }\textbf {\bibinfo {volume} {05}},\
  \bibinfo {pages} {109} (\bibinfo {year} {2022})},\ \Eprint
  {http://arxiv.org/abs/2202.11002} {arXiv:2202.11002 [hep-ph]} \BibitemShut
  {NoStop}%
\bibitem [{\citenamefont {Binh}\ \emph {et~al.}(2021)\citenamefont {Binh},
  \citenamefont {Hue}, \citenamefont {Binh}, \citenamefont {Soa},\ and\
  \citenamefont {Long}}]{Binh:2021iww}%
  \BibitemOpen
  \bibfield  {author} {\bibinfo {author} {\bibfnamefont {D.~T.}\ \bibnamefont
  {Binh}}, \bibinfo {author} {\bibfnamefont {L.~T.}\ \bibnamefont {Hue}},
  \bibinfo {author} {\bibfnamefont {V.~H.}\ \bibnamefont {Binh}}, \bibinfo
  {author} {\bibfnamefont {D.~V.}\ \bibnamefont {Soa}}, \ and\ \bibinfo
  {author} {\bibfnamefont {H.~N.}\ \bibnamefont {Long}},\ }\href@noop {} {\
  (\bibinfo {year} {2021})},\ \Eprint {http://arxiv.org/abs/2109.08118}
  {arXiv:2109.08118 [hep-ph]} \BibitemShut {NoStop}%
\bibitem [{\citenamefont {Bertuzzo}\ \emph {et~al.}(2022)\citenamefont
  {Bertuzzo}, \citenamefont {Grilli~di Cortona},\ and\ \citenamefont
  {Ramos}}]{Bertuzzo:2021opb}%
  \BibitemOpen
  \bibfield  {author} {\bibinfo {author} {\bibfnamefont {E.}~\bibnamefont
  {Bertuzzo}}, \bibinfo {author} {\bibfnamefont {G.}~\bibnamefont {Grilli~di
  Cortona}}, \ and\ \bibinfo {author} {\bibfnamefont {L.~M.~D.}\ \bibnamefont
  {Ramos}},\ }\href {\doibase 10.1007/JHEP06(2022)075} {\bibfield  {journal}
  {\bibinfo  {journal} {JHEP}\ }\textbf {\bibinfo {volume} {06}},\ \bibinfo
  {pages} {075} (\bibinfo {year} {2022})},\ \Eprint
  {http://arxiv.org/abs/2112.04020} {arXiv:2112.04020 [hep-ph]} \BibitemShut
  {NoStop}%
\bibitem [{\citenamefont {Chakraborty}\ \emph {et~al.}(2022)\citenamefont
  {Chakraborty}, \citenamefont {Das}, \citenamefont {Goswami},\ and\
  \citenamefont {Roy}}]{Chakraborty:2021apc}%
  \BibitemOpen
  \bibfield  {author} {\bibinfo {author} {\bibfnamefont {K.}~\bibnamefont
  {Chakraborty}}, \bibinfo {author} {\bibfnamefont {A.}~\bibnamefont {Das}},
  \bibinfo {author} {\bibfnamefont {S.}~\bibnamefont {Goswami}}, \ and\
  \bibinfo {author} {\bibfnamefont {S.}~\bibnamefont {Roy}},\ }\href {\doibase
  10.1007/JHEP04(2022)008} {\bibfield  {journal} {\bibinfo  {journal} {JHEP}\
  }\textbf {\bibinfo {volume} {04}},\ \bibinfo {pages} {008} (\bibinfo {year}
  {2022})},\ \Eprint {http://arxiv.org/abs/2111.08767} {arXiv:2111.08767
  [hep-ph]} \BibitemShut {NoStop}%
\bibitem [{\citenamefont {De~Romeri}\ \emph {et~al.}(2022)\citenamefont
  {De~Romeri}, \citenamefont {Miranda}, \citenamefont {Papoulias},
  \citenamefont {Sanchez~Garcia}, \citenamefont {T\'ortola},\ and\
  \citenamefont {Valle}}]{DeRomeri:2022twg}%
  \BibitemOpen
  \bibfield  {author} {\bibinfo {author} {\bibfnamefont {V.}~\bibnamefont
  {De~Romeri}}, \bibinfo {author} {\bibfnamefont {O.~G.}\ \bibnamefont
  {Miranda}}, \bibinfo {author} {\bibfnamefont {D.~K.}\ \bibnamefont
  {Papoulias}}, \bibinfo {author} {\bibfnamefont {G.}~\bibnamefont
  {Sanchez~Garcia}}, \bibinfo {author} {\bibfnamefont {M.}~\bibnamefont
  {T\'ortola}}, \ and\ \bibinfo {author} {\bibfnamefont {J.~W.~F.}\
  \bibnamefont {Valle}},\ }\href@noop {} {\  (\bibinfo {year} {2022})},\
  \Eprint {http://arxiv.org/abs/2211.11905} {arXiv:2211.11905 [hep-ph]}
  \BibitemShut {NoStop}%
\bibitem [{\citenamefont {Barranco}\ \emph {et~al.}(2005)\citenamefont
  {Barranco}, \citenamefont {Miranda},\ and\ \citenamefont
  {Rashba}}]{Barranco:2005yy}%
  \BibitemOpen
  \bibfield  {author} {\bibinfo {author} {\bibfnamefont {J.}~\bibnamefont
  {Barranco}}, \bibinfo {author} {\bibfnamefont {O.~G.}\ \bibnamefont
  {Miranda}}, \ and\ \bibinfo {author} {\bibfnamefont {T.~I.}\ \bibnamefont
  {Rashba}},\ }\href {\doibase 10.1088/1126-6708/2005/12/021} {\bibfield
  {journal} {\bibinfo  {journal} {JHEP}\ }\textbf {\bibinfo {volume} {12}},\
  \bibinfo {pages} {021} (\bibinfo {year} {2005})},\ \Eprint
  {http://arxiv.org/abs/hep-ph/0508299} {arXiv:hep-ph/0508299} \BibitemShut
  {NoStop}%
\bibitem [{\citenamefont {Scholberg}(2006)}]{Scholberg:2005qs}%
  \BibitemOpen
  \bibfield  {author} {\bibinfo {author} {\bibfnamefont {K.}~\bibnamefont
  {Scholberg}},\ }\href {\doibase 10.1103/PhysRevD.73.033005} {\bibfield
  {journal} {\bibinfo  {journal} {Phys. Rev. D}\ }\textbf {\bibinfo {volume}
  {73}},\ \bibinfo {pages} {033005} (\bibinfo {year} {2006})},\ \Eprint
  {http://arxiv.org/abs/hep-ex/0511042} {arXiv:hep-ex/0511042} \BibitemShut
  {NoStop}%
\bibitem [{\citenamefont {Lindner}\ \emph {et~al.}(2017)\citenamefont
  {Lindner}, \citenamefont {Rodejohann},\ and\ \citenamefont
  {Xu}}]{Lindner:2016wff}%
  \BibitemOpen
  \bibfield  {author} {\bibinfo {author} {\bibfnamefont {M.}~\bibnamefont
  {Lindner}}, \bibinfo {author} {\bibfnamefont {W.}~\bibnamefont {Rodejohann}},
  \ and\ \bibinfo {author} {\bibfnamefont {X.-J.}\ \bibnamefont {Xu}},\ }\href
  {\doibase 10.1007/JHEP03(2017)097} {\bibfield  {journal} {\bibinfo  {journal}
  {JHEP}\ }\textbf {\bibinfo {volume} {03}},\ \bibinfo {pages} {097} (\bibinfo
  {year} {2017})},\ \Eprint {http://arxiv.org/abs/1612.04150} {arXiv:1612.04150
  [hep-ph]} \BibitemShut {NoStop}%
\bibitem [{\citenamefont {Liao}\ and\ \citenamefont
  {Marfatia}(2017)}]{Liao:2017uzy}%
  \BibitemOpen
  \bibfield  {author} {\bibinfo {author} {\bibfnamefont {J.}~\bibnamefont
  {Liao}}\ and\ \bibinfo {author} {\bibfnamefont {D.}~\bibnamefont
  {Marfatia}},\ }\href {\doibase 10.1016/j.physletb.2017.10.046} {\bibfield
  {journal} {\bibinfo  {journal} {Phys. Lett. B}\ }\textbf {\bibinfo {volume}
  {775}},\ \bibinfo {pages} {54} (\bibinfo {year} {2017})},\ \Eprint
  {http://arxiv.org/abs/1708.04255} {arXiv:1708.04255 [hep-ph]} \BibitemShut
  {NoStop}%
\bibitem [{\citenamefont {Giunti}(2020)}]{Giunti:2019xpr}%
  \BibitemOpen
  \bibfield  {author} {\bibinfo {author} {\bibfnamefont {C.}~\bibnamefont
  {Giunti}},\ }\href {\doibase 10.1103/PhysRevD.101.035039} {\bibfield
  {journal} {\bibinfo  {journal} {Phys. Rev. D}\ }\textbf {\bibinfo {volume}
  {101}},\ \bibinfo {pages} {035039} (\bibinfo {year} {2020})},\ \Eprint
  {http://arxiv.org/abs/1909.00466} {arXiv:1909.00466 [hep-ph]} \BibitemShut
  {NoStop}%
\bibitem [{\citenamefont {Denton}\ and\ \citenamefont
  {Gehrlein}(2021)}]{Denton:2020hop}%
  \BibitemOpen
  \bibfield  {author} {\bibinfo {author} {\bibfnamefont {P.~B.}\ \bibnamefont
  {Denton}}\ and\ \bibinfo {author} {\bibfnamefont {J.}~\bibnamefont
  {Gehrlein}},\ }\href {\doibase 10.1007/JHEP04(2021)266} {\bibfield  {journal}
  {\bibinfo  {journal} {JHEP}\ }\textbf {\bibinfo {volume} {04}},\ \bibinfo
  {pages} {266} (\bibinfo {year} {2021})},\ \Eprint
  {http://arxiv.org/abs/2008.06062} {arXiv:2008.06062 [hep-ph]} \BibitemShut
  {NoStop}%
\bibitem [{\citenamefont {Khan}\ \emph {et~al.}(2021)\citenamefont {Khan},
  \citenamefont {McKay},\ and\ \citenamefont {Rodejohann}}]{Khan:2021wzy}%
  \BibitemOpen
  \bibfield  {author} {\bibinfo {author} {\bibfnamefont {A.~N.}\ \bibnamefont
  {Khan}}, \bibinfo {author} {\bibfnamefont {D.~W.}\ \bibnamefont {McKay}}, \
  and\ \bibinfo {author} {\bibfnamefont {W.}~\bibnamefont {Rodejohann}},\
  }\href {\doibase 10.1103/PhysRevD.104.015019} {\bibfield  {journal} {\bibinfo
   {journal} {Phys. Rev. D}\ }\textbf {\bibinfo {volume} {104}},\ \bibinfo
  {pages} {015019} (\bibinfo {year} {2021})},\ \Eprint
  {http://arxiv.org/abs/2104.00425} {arXiv:2104.00425 [hep-ph]} \BibitemShut
  {NoStop}%
\bibitem [{\citenamefont {Coloma}\ \emph {et~al.}(2022)\citenamefont {Coloma},
  \citenamefont {Esteban}, \citenamefont {Gonzalez-Garcia}, \citenamefont
  {Larizgoitia}, \citenamefont {Monrabal},\ and\ \citenamefont
  {Palomares-Ruiz}}]{Coloma:2022avw}%
  \BibitemOpen
  \bibfield  {author} {\bibinfo {author} {\bibfnamefont {P.}~\bibnamefont
  {Coloma}}, \bibinfo {author} {\bibfnamefont {I.}~\bibnamefont {Esteban}},
  \bibinfo {author} {\bibfnamefont {M.~C.}\ \bibnamefont {Gonzalez-Garcia}},
  \bibinfo {author} {\bibfnamefont {L.}~\bibnamefont {Larizgoitia}}, \bibinfo
  {author} {\bibfnamefont {F.}~\bibnamefont {Monrabal}}, \ and\ \bibinfo
  {author} {\bibfnamefont {S.}~\bibnamefont {Palomares-Ruiz}},\ }\href
  {\doibase 10.1007/JHEP05(2022)037} {\bibfield  {journal} {\bibinfo  {journal}
  {JHEP}\ }\textbf {\bibinfo {volume} {05}},\ \bibinfo {pages} {037} (\bibinfo
  {year} {2022})},\ \Eprint {http://arxiv.org/abs/2202.10829} {arXiv:2202.10829
  [hep-ph]} \BibitemShut {NoStop}%
\bibitem [{\citenamefont {Chatterjee}\ \emph {et~al.}(2022)\citenamefont
  {Chatterjee}, \citenamefont {Lavignac}, \citenamefont {Miranda},\ and\
  \citenamefont {Sanchez~Garcia}}]{Chatterjee:2022mmu}%
  \BibitemOpen
  \bibfield  {author} {\bibinfo {author} {\bibfnamefont {S.~S.}\ \bibnamefont
  {Chatterjee}}, \bibinfo {author} {\bibfnamefont {S.}~\bibnamefont
  {Lavignac}}, \bibinfo {author} {\bibfnamefont {O.~G.}\ \bibnamefont
  {Miranda}}, \ and\ \bibinfo {author} {\bibfnamefont {G.}~\bibnamefont
  {Sanchez~Garcia}},\ }\href@noop {} {\  (\bibinfo {year} {2022})},\ \Eprint
  {http://arxiv.org/abs/2208.11771} {arXiv:2208.11771 [hep-ph]} \BibitemShut
  {NoStop}%
\bibitem [{\citenamefont {Dutta}\ \emph
  {et~al.}(2016{\natexlab{b}})\citenamefont {Dutta}, \citenamefont {Gao},
  \citenamefont {Mahapatra}, \citenamefont {Mirabolfathi}, \citenamefont
  {Strigari},\ and\ \citenamefont {Walker}}]{Dutta:2015nlo}%
  \BibitemOpen
  \bibfield  {author} {\bibinfo {author} {\bibfnamefont {B.}~\bibnamefont
  {Dutta}}, \bibinfo {author} {\bibfnamefont {Y.}~\bibnamefont {Gao}}, \bibinfo
  {author} {\bibfnamefont {R.}~\bibnamefont {Mahapatra}}, \bibinfo {author}
  {\bibfnamefont {N.}~\bibnamefont {Mirabolfathi}}, \bibinfo {author}
  {\bibfnamefont {L.~E.}\ \bibnamefont {Strigari}}, \ and\ \bibinfo {author}
  {\bibfnamefont {J.~W.}\ \bibnamefont {Walker}},\ }\href {\doibase
  10.1103/PhysRevD.94.093002} {\bibfield  {journal} {\bibinfo  {journal} {Phys.
  Rev. D}\ }\textbf {\bibinfo {volume} {94}},\ \bibinfo {pages} {093002}
  (\bibinfo {year} {2016}{\natexlab{b}})},\ \Eprint
  {http://arxiv.org/abs/1511.02834} {arXiv:1511.02834 [hep-ph]} \BibitemShut
  {NoStop}%
\bibitem [{\citenamefont {Flores}\ \emph {et~al.}(2022)\citenamefont {Flores},
  \citenamefont {Nath},\ and\ \citenamefont {Peinado}}]{Flores:2021kzl}%
  \BibitemOpen
  \bibfield  {author} {\bibinfo {author} {\bibfnamefont {L.~J.}\ \bibnamefont
  {Flores}}, \bibinfo {author} {\bibfnamefont {N.}~\bibnamefont {Nath}}, \ and\
  \bibinfo {author} {\bibfnamefont {E.}~\bibnamefont {Peinado}},\ }\href
  {\doibase 10.1103/PhysRevD.105.055010} {\bibfield  {journal} {\bibinfo
  {journal} {Phys. Rev. D}\ }\textbf {\bibinfo {volume} {105}},\ \bibinfo
  {pages} {055010} (\bibinfo {year} {2022})},\ \Eprint
  {http://arxiv.org/abs/2112.05103} {arXiv:2112.05103 [hep-ph]} \BibitemShut
  {NoStop}%
\bibitem [{\citenamefont {Miranda}\ \emph
  {et~al.}(2019{\natexlab{a}})\citenamefont {Miranda}, \citenamefont
  {Papoulias}, \citenamefont {T\'ortola},\ and\ \citenamefont
  {Valle}}]{Miranda:2019wdy}%
  \BibitemOpen
  \bibfield  {author} {\bibinfo {author} {\bibfnamefont {O.~G.}\ \bibnamefont
  {Miranda}}, \bibinfo {author} {\bibfnamefont {D.~K.}\ \bibnamefont
  {Papoulias}}, \bibinfo {author} {\bibfnamefont {M.}~\bibnamefont
  {T\'ortola}}, \ and\ \bibinfo {author} {\bibfnamefont {J.~W.~F.}\
  \bibnamefont {Valle}},\ }\href {\doibase 10.1007/JHEP07(2019)103} {\bibfield
  {journal} {\bibinfo  {journal} {JHEP}\ }\textbf {\bibinfo {volume} {07}},\
  \bibinfo {pages} {103} (\bibinfo {year} {2019}{\natexlab{a}})},\ \Eprint
  {http://arxiv.org/abs/1905.03750} {arXiv:1905.03750 [hep-ph]} \BibitemShut
  {NoStop}%
\bibitem [{\citenamefont {Miranda}\ \emph
  {et~al.}(2020{\natexlab{b}})\citenamefont {Miranda}, \citenamefont
  {Papoulias}, \citenamefont {Sanders}, \citenamefont {T\'ortola},\ and\
  \citenamefont {Valle}}]{Miranda:2020syh}%
  \BibitemOpen
  \bibfield  {author} {\bibinfo {author} {\bibfnamefont {O.~G.}\ \bibnamefont
  {Miranda}}, \bibinfo {author} {\bibfnamefont {D.~K.}\ \bibnamefont
  {Papoulias}}, \bibinfo {author} {\bibfnamefont {O.}~\bibnamefont {Sanders}},
  \bibinfo {author} {\bibfnamefont {M.}~\bibnamefont {T\'ortola}}, \ and\
  \bibinfo {author} {\bibfnamefont {J.~W.~F.}\ \bibnamefont {Valle}},\ }\href
  {\doibase 10.1103/PhysRevD.102.113014} {\bibfield  {journal} {\bibinfo
  {journal} {Phys. Rev. D}\ }\textbf {\bibinfo {volume} {102}},\ \bibinfo
  {pages} {113014} (\bibinfo {year} {2020}{\natexlab{b}})},\ \Eprint
  {http://arxiv.org/abs/2008.02759} {arXiv:2008.02759 [hep-ph]} \BibitemShut
  {NoStop}%
\bibitem [{\citenamefont {Kosmas}\ \emph {et~al.}(2017)\citenamefont {Kosmas},
  \citenamefont {Papoulias}, \citenamefont {Tortola},\ and\ \citenamefont
  {Valle}}]{Kosmas:2017zbh}%
  \BibitemOpen
  \bibfield  {author} {\bibinfo {author} {\bibfnamefont {T.~S.}\ \bibnamefont
  {Kosmas}}, \bibinfo {author} {\bibfnamefont {D.~K.}\ \bibnamefont
  {Papoulias}}, \bibinfo {author} {\bibfnamefont {M.}~\bibnamefont {Tortola}},
  \ and\ \bibinfo {author} {\bibfnamefont {J.~W.~F.}\ \bibnamefont {Valle}},\
  }\href {\doibase 10.1103/PhysRevD.96.063013} {\bibfield  {journal} {\bibinfo
  {journal} {Phys. Rev. D}\ }\textbf {\bibinfo {volume} {96}},\ \bibinfo
  {pages} {063013} (\bibinfo {year} {2017})},\ \Eprint
  {http://arxiv.org/abs/1703.00054} {arXiv:1703.00054 [hep-ph]} \BibitemShut
  {NoStop}%
\bibitem [{\citenamefont {Miranda}\ \emph
  {et~al.}(2019{\natexlab{b}})\citenamefont {Miranda}, \citenamefont
  {Sanchez~Garcia},\ and\ \citenamefont {Sanders}}]{Miranda:2019skf}%
  \BibitemOpen
  \bibfield  {author} {\bibinfo {author} {\bibfnamefont {O.~G.}\ \bibnamefont
  {Miranda}}, \bibinfo {author} {\bibfnamefont {G.}~\bibnamefont
  {Sanchez~Garcia}}, \ and\ \bibinfo {author} {\bibfnamefont {O.}~\bibnamefont
  {Sanders}},\ }\href {\doibase 10.1155/2019/3902819} {\bibfield  {journal}
  {\bibinfo  {journal} {Adv. High Energy Phys.}\ }\textbf {\bibinfo {volume}
  {2019}},\ \bibinfo {pages} {3902819} (\bibinfo {year}
  {2019}{\natexlab{b}})},\ \Eprint {http://arxiv.org/abs/1902.09036}
  {arXiv:1902.09036 [hep-ph]} \BibitemShut {NoStop}%
\bibitem [{\citenamefont {Dasgupta}\ \emph {et~al.}(2021)\citenamefont
  {Dasgupta}, \citenamefont {Kang},\ and\ \citenamefont
  {Kim}}]{Dasgupta:2021fpn}%
  \BibitemOpen
  \bibfield  {author} {\bibinfo {author} {\bibfnamefont {A.}~\bibnamefont
  {Dasgupta}}, \bibinfo {author} {\bibfnamefont {S.~K.}\ \bibnamefont {Kang}},
  \ and\ \bibinfo {author} {\bibfnamefont {J.~E.}\ \bibnamefont {Kim}},\ }\href
  {\doibase 10.1007/JHEP11(2021)120} {\bibfield  {journal} {\bibinfo  {journal}
  {JHEP}\ }\textbf {\bibinfo {volume} {11}},\ \bibinfo {pages} {120} (\bibinfo
  {year} {2021})},\ \Eprint {http://arxiv.org/abs/2108.12998} {arXiv:2108.12998
  [hep-ph]} \BibitemShut {NoStop}%
\bibitem [{\citenamefont {Bolton}\ \emph {et~al.}(2022)\citenamefont {Bolton},
  \citenamefont {Deppisch}, \citenamefont {Fridell}, \citenamefont {Harz},
  \citenamefont {Hati},\ and\ \citenamefont {Kulkarni}}]{Bolton:2021pey}%
  \BibitemOpen
  \bibfield  {author} {\bibinfo {author} {\bibfnamefont {P.~D.}\ \bibnamefont
  {Bolton}}, \bibinfo {author} {\bibfnamefont {F.~F.}\ \bibnamefont
  {Deppisch}}, \bibinfo {author} {\bibfnamefont {K.}~\bibnamefont {Fridell}},
  \bibinfo {author} {\bibfnamefont {J.}~\bibnamefont {Harz}}, \bibinfo {author}
  {\bibfnamefont {C.}~\bibnamefont {Hati}}, \ and\ \bibinfo {author}
  {\bibfnamefont {S.}~\bibnamefont {Kulkarni}},\ }\href {\doibase
  10.1103/PhysRevD.106.035036} {\bibfield  {journal} {\bibinfo  {journal}
  {Phys. Rev. D}\ }\textbf {\bibinfo {volume} {106}},\ \bibinfo {pages}
  {035036} (\bibinfo {year} {2022})},\ \Eprint
  {http://arxiv.org/abs/2110.02233} {arXiv:2110.02233 [hep-ph]} \BibitemShut
  {NoStop}%
\bibitem [{\citenamefont {Cadeddu}\ \emph
  {et~al.}(2020{\natexlab{b}})\citenamefont {Cadeddu}, \citenamefont {Dordei},
  \citenamefont {Giunti}, \citenamefont {Li}, \citenamefont {Picciau},\ and\
  \citenamefont {Zhang}}]{Cadeddu:2020lky}%
  \BibitemOpen
  \bibfield  {author} {\bibinfo {author} {\bibfnamefont {M.}~\bibnamefont
  {Cadeddu}}, \bibinfo {author} {\bibfnamefont {F.}~\bibnamefont {Dordei}},
  \bibinfo {author} {\bibfnamefont {C.}~\bibnamefont {Giunti}}, \bibinfo
  {author} {\bibfnamefont {Y.~F.}\ \bibnamefont {Li}}, \bibinfo {author}
  {\bibfnamefont {E.}~\bibnamefont {Picciau}}, \ and\ \bibinfo {author}
  {\bibfnamefont {Y.~Y.}\ \bibnamefont {Zhang}},\ }\href {\doibase
  10.1103/PhysRevD.102.015030} {\bibfield  {journal} {\bibinfo  {journal}
  {Phys. Rev. D}\ }\textbf {\bibinfo {volume} {102}},\ \bibinfo {pages}
  {015030} (\bibinfo {year} {2020}{\natexlab{b}})},\ \Eprint
  {http://arxiv.org/abs/2005.01645} {arXiv:2005.01645 [hep-ph]} \BibitemShut
  {NoStop}%
\bibitem [{\citenamefont {Brdar}\ \emph {et~al.}(2018)\citenamefont {Brdar},
  \citenamefont {Rodejohann},\ and\ \citenamefont {Xu}}]{Brdar:2018qqj}%
  \BibitemOpen
  \bibfield  {author} {\bibinfo {author} {\bibfnamefont {V.}~\bibnamefont
  {Brdar}}, \bibinfo {author} {\bibfnamefont {W.}~\bibnamefont {Rodejohann}}, \
  and\ \bibinfo {author} {\bibfnamefont {X.-J.}\ \bibnamefont {Xu}},\ }\href
  {\doibase 10.1007/JHEP12(2018)024} {\bibfield  {journal} {\bibinfo  {journal}
  {JHEP}\ }\textbf {\bibinfo {volume} {12}},\ \bibinfo {pages} {024} (\bibinfo
  {year} {2018})},\ \Eprint {http://arxiv.org/abs/1810.03626} {arXiv:1810.03626
  [hep-ph]} \BibitemShut {NoStop}%
\bibitem [{\citenamefont {Dror}\ \emph
  {et~al.}(2020{\natexlab{a}})\citenamefont {Dror}, \citenamefont {Elor},\ and\
  \citenamefont {Mcgehee}}]{Dror:2019onn}%
  \BibitemOpen
  \bibfield  {author} {\bibinfo {author} {\bibfnamefont {J.~A.}\ \bibnamefont
  {Dror}}, \bibinfo {author} {\bibfnamefont {G.}~\bibnamefont {Elor}}, \ and\
  \bibinfo {author} {\bibfnamefont {R.}~\bibnamefont {Mcgehee}},\ }\href
  {\doibase 10.1103/PhysRevLett.124.181301} {\bibfield  {journal} {\bibinfo
  {journal} {Phys. Rev. Lett.}\ }\textbf {\bibinfo {volume} {124}},\ \bibinfo
  {pages} {18} (\bibinfo {year} {2020}{\natexlab{a}})},\ \Eprint
  {http://arxiv.org/abs/1905.12635} {arXiv:1905.12635 [hep-ph]} \BibitemShut
  {NoStop}%
\bibitem [{\citenamefont {Dror}\ \emph
  {et~al.}(2020{\natexlab{b}})\citenamefont {Dror}, \citenamefont {Elor},\ and\
  \citenamefont {Mcgehee}}]{Dror:2019dib}%
  \BibitemOpen
  \bibfield  {author} {\bibinfo {author} {\bibfnamefont {J.~A.}\ \bibnamefont
  {Dror}}, \bibinfo {author} {\bibfnamefont {G.}~\bibnamefont {Elor}}, \ and\
  \bibinfo {author} {\bibfnamefont {R.}~\bibnamefont {Mcgehee}},\ }\href
  {\doibase 10.1007/JHEP02(2020)134} {\bibfield  {journal} {\bibinfo  {journal}
  {JHEP}\ }\textbf {\bibinfo {volume} {02}},\ \bibinfo {pages} {134} (\bibinfo
  {year} {2020}{\natexlab{b}})},\ \Eprint {http://arxiv.org/abs/1908.10861}
  {arXiv:1908.10861 [hep-ph]} \BibitemShut {NoStop}%
\bibitem [{\citenamefont {Farzan}\ \emph {et~al.}(2018)\citenamefont {Farzan},
  \citenamefont {Lindner}, \citenamefont {Rodejohann},\ and\ \citenamefont
  {Xu}}]{Farzan:2018gtr}%
  \BibitemOpen
  \bibfield  {author} {\bibinfo {author} {\bibfnamefont {Y.}~\bibnamefont
  {Farzan}}, \bibinfo {author} {\bibfnamefont {M.}~\bibnamefont {Lindner}},
  \bibinfo {author} {\bibfnamefont {W.}~\bibnamefont {Rodejohann}}, \ and\
  \bibinfo {author} {\bibfnamefont {X.-J.}\ \bibnamefont {Xu}},\ }\href
  {\doibase 10.1007/JHEP05(2018)066} {\bibfield  {journal} {\bibinfo  {journal}
  {JHEP}\ }\textbf {\bibinfo {volume} {05}},\ \bibinfo {pages} {066} (\bibinfo
  {year} {2018})},\ \Eprint {http://arxiv.org/abs/1802.05171} {arXiv:1802.05171
  [hep-ph]} \BibitemShut {NoStop}%
\bibitem [{\citenamefont {Dey}\ and\ \citenamefont
  {Mohanty}(2016)}]{Dey:2015eaa}%
  \BibitemOpen
  \bibfield  {author} {\bibinfo {author} {\bibfnamefont {U.~K.}\ \bibnamefont
  {Dey}}\ and\ \bibinfo {author} {\bibfnamefont {S.}~\bibnamefont {Mohanty}},\
  }\href {\doibase 10.1007/JHEP04(2016)187} {\bibfield  {journal} {\bibinfo
  {journal} {JHEP}\ }\textbf {\bibinfo {volume} {04}},\ \bibinfo {pages} {187}
  (\bibinfo {year} {2016})},\ \Eprint {http://arxiv.org/abs/1505.01037}
  {arXiv:1505.01037 [hep-ph]} \BibitemShut {NoStop}%
\bibitem [{\citenamefont {Valencia}\ and\ \citenamefont
  {Willenbrock}(1994)}]{Valencia:1994cj}%
  \BibitemOpen
  \bibfield  {author} {\bibinfo {author} {\bibfnamefont {G.}~\bibnamefont
  {Valencia}}\ and\ \bibinfo {author} {\bibfnamefont {S.}~\bibnamefont
  {Willenbrock}},\ }\href {\doibase 10.1103/PhysRevD.50.6843} {\bibfield
  {journal} {\bibinfo  {journal} {Phys. Rev. D}\ }\textbf {\bibinfo {volume}
  {50}},\ \bibinfo {pages} {6843} (\bibinfo {year} {1994})},\ \Eprint
  {http://arxiv.org/abs/hep-ph/9409201} {arXiv:hep-ph/9409201} \BibitemShut
  {NoStop}%
\bibitem [{\citenamefont {Smirnov}(2007)}]{Smirnov:2007hv}%
  \BibitemOpen
  \bibfield  {author} {\bibinfo {author} {\bibfnamefont {A.~D.}\ \bibnamefont
  {Smirnov}},\ }\href {\doibase 10.1142/S0217732307024401} {\bibfield
  {journal} {\bibinfo  {journal} {Mod. Phys. Lett. A}\ }\textbf {\bibinfo
  {volume} {22}},\ \bibinfo {pages} {2353} (\bibinfo {year} {2007})},\ \Eprint
  {http://arxiv.org/abs/0705.0308} {arXiv:0705.0308 [hep-ph]} \BibitemShut
  {NoStop}%
\bibitem [{\citenamefont {Georgi}\ and\ \citenamefont
  {Glashow}(1974)}]{PhysRevLett.32.438}%
  \BibitemOpen
  \bibfield  {author} {\bibinfo {author} {\bibfnamefont {H.}~\bibnamefont
  {Georgi}}\ and\ \bibinfo {author} {\bibfnamefont {S.~L.}\ \bibnamefont
  {Glashow}},\ }\href {\doibase 10.1103/PhysRevLett.32.438} {\bibfield
  {journal} {\bibinfo  {journal} {Phys. Rev. Lett.}\ }\textbf {\bibinfo
  {volume} {32}},\ \bibinfo {pages} {438} (\bibinfo {year} {1974})}\BibitemShut
  {NoStop}%
\bibitem [{\citenamefont {Fritzsch}\ and\ \citenamefont
  {Minkowski}(1975)}]{FRITZSCH1975193}%
  \BibitemOpen
  \bibfield  {author} {\bibinfo {author} {\bibfnamefont {H.}~\bibnamefont
  {Fritzsch}}\ and\ \bibinfo {author} {\bibfnamefont {P.}~\bibnamefont
  {Minkowski}},\ }\href {\doibase https://doi.org/10.1016/0003-4916(75)90211-0}
  {\bibfield  {journal} {\bibinfo  {journal} {Annals of Physics}\ }\textbf
  {\bibinfo {volume} {93}},\ \bibinfo {pages} {193} (\bibinfo {year}
  {1975})}\BibitemShut {NoStop}%
\bibitem [{\citenamefont {Preda}\ \emph {et~al.}(2022)\citenamefont {Preda},
  \citenamefont {Senjanovic},\ and\ \citenamefont
  {Zantedeschi}}]{Preda:2022izo}%
  \BibitemOpen
  \bibfield  {author} {\bibinfo {author} {\bibfnamefont {A.}~\bibnamefont
  {Preda}}, \bibinfo {author} {\bibfnamefont {G.}~\bibnamefont {Senjanovic}}, \
  and\ \bibinfo {author} {\bibfnamefont {M.}~\bibnamefont {Zantedeschi}},\
  }\href@noop {} {\  (\bibinfo {year} {2022})},\ \Eprint
  {http://arxiv.org/abs/2201.02785} {arXiv:2201.02785 [hep-ph]} \BibitemShut
  {NoStop}%
\bibitem [{\citenamefont {Pati}\ and\ \citenamefont
  {Salam}(1974)}]{PhysRevD.10.275}%
  \BibitemOpen
  \bibfield  {author} {\bibinfo {author} {\bibfnamefont {J.~C.}\ \bibnamefont
  {Pati}}\ and\ \bibinfo {author} {\bibfnamefont {A.}~\bibnamefont {Salam}},\
  }\href {\doibase 10.1103/PhysRevD.10.275} {\bibfield  {journal} {\bibinfo
  {journal} {Phys. Rev. D}\ }\textbf {\bibinfo {volume} {10}},\ \bibinfo
  {pages} {275} (\bibinfo {year} {1974})}\BibitemShut {NoStop}%
\bibitem [{\citenamefont {Farhi}\ and\ \citenamefont
  {Susskind}(1981)}]{FARHI1981277}%
  \BibitemOpen
  \bibfield  {author} {\bibinfo {author} {\bibfnamefont {E.}~\bibnamefont
  {Farhi}}\ and\ \bibinfo {author} {\bibfnamefont {L.}~\bibnamefont
  {Susskind}},\ }\href {\doibase https://doi.org/10.1016/0370-1573(81)90173-3}
  {\bibfield  {journal} {\bibinfo  {journal} {Physics Reports}\ }\textbf
  {\bibinfo {volume} {74}},\ \bibinfo {pages} {277} (\bibinfo {year}
  {1981})}\BibitemShut {NoStop}%
\bibitem [{\citenamefont {Gabriel}\ \emph {et~al.}(1994)\citenamefont
  {Gabriel}, \citenamefont {Groom}, \citenamefont {Job}, \citenamefont
  {Mokhov},\ and\ \citenamefont {Stevenson}}]{GABRIEL1994336}%
  \BibitemOpen
  \bibfield  {author} {\bibinfo {author} {\bibfnamefont {T.}~\bibnamefont
  {Gabriel}}, \bibinfo {author} {\bibfnamefont {D.}~\bibnamefont {Groom}},
  \bibinfo {author} {\bibfnamefont {P.}~\bibnamefont {Job}}, \bibinfo {author}
  {\bibfnamefont {N.}~\bibnamefont {Mokhov}}, \ and\ \bibinfo {author}
  {\bibfnamefont {G.}~\bibnamefont {Stevenson}},\ }\href {\doibase
  https://doi.org/10.1016/0168-9002(94)91317-X} {\bibfield  {journal} {\bibinfo
   {journal} {Nuclear Instruments and Methods in Physics Research Section A:
  Accelerators, Spectrometers, Detectors and Associated Equipment}\ }\textbf
  {\bibinfo {volume} {338}},\ \bibinfo {pages} {336} (\bibinfo {year}
  {1994})}\BibitemShut {NoStop}%
\bibitem [{\citenamefont {Andersen}\ \emph {et~al.}(2011)\citenamefont
  {Andersen} \emph {et~al.}}]{Andersen:2011yj}%
  \BibitemOpen
  \bibfield  {author} {\bibinfo {author} {\bibfnamefont {J.~R.}\ \bibnamefont
  {Andersen}} \emph {et~al.},\ }\href {\doibase 10.1140/epjp/i2011-11081-1}
  {\bibfield  {journal} {\bibinfo  {journal} {Eur. Phys. J. Plus}\ }\textbf
  {\bibinfo {volume} {126}},\ \bibinfo {pages} {81} (\bibinfo {year} {2011})},\
  \Eprint {http://arxiv.org/abs/1104.1255} {arXiv:1104.1255 [hep-ph]}
  \BibitemShut {NoStop}%
\bibitem [{\citenamefont {Buchmüller}\ \emph {et~al.}(1987)\citenamefont
  {Buchmüller}, \citenamefont {Rückl},\ and\ \citenamefont
  {Wyler}}]{BUCHMULLER1987442}%
  \BibitemOpen
  \bibfield  {author} {\bibinfo {author} {\bibfnamefont {W.}~\bibnamefont
  {Buchmüller}}, \bibinfo {author} {\bibfnamefont {R.}~\bibnamefont {Rückl}},
  \ and\ \bibinfo {author} {\bibfnamefont {D.}~\bibnamefont {Wyler}},\ }\href
  {\doibase https://doi.org/10.1016/0370-2693(87)90637-X} {\bibfield  {journal}
  {\bibinfo  {journal} {Physics Letters B}\ }\textbf {\bibinfo {volume}
  {191}},\ \bibinfo {pages} {442} (\bibinfo {year} {1987})}\BibitemShut
  {NoStop}%
\bibitem [{\citenamefont {Belyaev}\ \emph {et~al.}(2005)\citenamefont
  {Belyaev}, \citenamefont {Leroy}, \citenamefont {Mehdiyev},\ and\
  \citenamefont {Pukhov}}]{Belyaev:2005ew}%
  \BibitemOpen
  \bibfield  {author} {\bibinfo {author} {\bibfnamefont {A.}~\bibnamefont
  {Belyaev}}, \bibinfo {author} {\bibfnamefont {C.}~\bibnamefont {Leroy}},
  \bibinfo {author} {\bibfnamefont {R.}~\bibnamefont {Mehdiyev}}, \ and\
  \bibinfo {author} {\bibfnamefont {A.}~\bibnamefont {Pukhov}},\ }\href
  {\doibase 10.1088/1126-6708/2005/09/005} {\bibfield  {journal} {\bibinfo
  {journal} {JHEP}\ }\textbf {\bibinfo {volume} {09}},\ \bibinfo {pages} {005}
  (\bibinfo {year} {2005})},\ \Eprint {http://arxiv.org/abs/hep-ph/0502067}
  {arXiv:hep-ph/0502067} \BibitemShut {NoStop}%
\bibitem [{\citenamefont {Dorsner}\ and\ \citenamefont
  {Fileviez~Perez}(2005)}]{Dorsner:2005fq}%
  \BibitemOpen
  \bibfield  {author} {\bibinfo {author} {\bibfnamefont {I.}~\bibnamefont
  {Dorsner}}\ and\ \bibinfo {author} {\bibfnamefont {P.}~\bibnamefont
  {Fileviez~Perez}},\ }\href {\doibase 10.1016/j.nuclphysb.2005.06.016}
  {\bibfield  {journal} {\bibinfo  {journal} {Nucl. Phys. B}\ }\textbf
  {\bibinfo {volume} {723}},\ \bibinfo {pages} {53} (\bibinfo {year} {2005})},\
  \Eprint {http://arxiv.org/abs/hep-ph/0504276} {arXiv:hep-ph/0504276}
  \BibitemShut {NoStop}%
\bibitem [{\citenamefont {Fileviez~Perez}\ and\ \citenamefont
  {Wise}(2013)}]{FileviezPerez:2013zmv}%
  \BibitemOpen
  \bibfield  {author} {\bibinfo {author} {\bibfnamefont {P.}~\bibnamefont
  {Fileviez~Perez}}\ and\ \bibinfo {author} {\bibfnamefont {M.~B.}\
  \bibnamefont {Wise}},\ }\href {\doibase 10.1103/PhysRevD.88.057703}
  {\bibfield  {journal} {\bibinfo  {journal} {Phys. Rev. D}\ }\textbf {\bibinfo
  {volume} {88}},\ \bibinfo {pages} {057703} (\bibinfo {year} {2013})},\
  \Eprint {http://arxiv.org/abs/1307.6213} {arXiv:1307.6213 [hep-ph]}
  \BibitemShut {NoStop}%
\bibitem [{\citenamefont {Crivellin}\ \emph
  {et~al.}(2021{\natexlab{a}})\citenamefont {Crivellin}, \citenamefont
  {Hoferichter}, \citenamefont {Kirk}, \citenamefont {Manzari},\ and\
  \citenamefont {Schnell}}]{Crivellin:2021bkd}%
  \BibitemOpen
  \bibfield  {author} {\bibinfo {author} {\bibfnamefont {A.}~\bibnamefont
  {Crivellin}}, \bibinfo {author} {\bibfnamefont {M.}~\bibnamefont
  {Hoferichter}}, \bibinfo {author} {\bibfnamefont {M.}~\bibnamefont {Kirk}},
  \bibinfo {author} {\bibfnamefont {C.~A.}\ \bibnamefont {Manzari}}, \ and\
  \bibinfo {author} {\bibfnamefont {L.}~\bibnamefont {Schnell}},\ }\href
  {\doibase 10.1007/JHEP10(2021)221} {\bibfield  {journal} {\bibinfo  {journal}
  {JHEP}\ }\textbf {\bibinfo {volume} {10}},\ \bibinfo {pages} {221} (\bibinfo
  {year} {2021}{\natexlab{a}})},\ \Eprint {http://arxiv.org/abs/2107.13569}
  {arXiv:2107.13569 [hep-ph]} \BibitemShut {NoStop}%
\bibitem [{\citenamefont {Davidson}\ \emph {et~al.}(1994)\citenamefont
  {Davidson}, \citenamefont {Bailey},\ and\ \citenamefont
  {Campbell}}]{Davidson:1993qk}%
  \BibitemOpen
  \bibfield  {author} {\bibinfo {author} {\bibfnamefont {S.}~\bibnamefont
  {Davidson}}, \bibinfo {author} {\bibfnamefont {D.~C.}\ \bibnamefont
  {Bailey}}, \ and\ \bibinfo {author} {\bibfnamefont {B.~A.}\ \bibnamefont
  {Campbell}},\ }\href {\doibase 10.1007/BF01552629} {\bibfield  {journal}
  {\bibinfo  {journal} {Z. Phys. C}\ }\textbf {\bibinfo {volume} {61}},\
  \bibinfo {pages} {613} (\bibinfo {year} {1994})},\ \Eprint
  {http://arxiv.org/abs/hep-ph/9309310} {arXiv:hep-ph/9309310} \BibitemShut
  {NoStop}%
\bibitem [{\citenamefont {Crivellin}\ \emph
  {et~al.}(2021{\natexlab{b}})\citenamefont {Crivellin}, \citenamefont
  {M\"uller},\ and\ \citenamefont {Schnell}}]{Crivellin:2021egp}%
  \BibitemOpen
  \bibfield  {author} {\bibinfo {author} {\bibfnamefont {A.}~\bibnamefont
  {Crivellin}}, \bibinfo {author} {\bibfnamefont {D.}~\bibnamefont {M\"uller}},
  \ and\ \bibinfo {author} {\bibfnamefont {L.}~\bibnamefont {Schnell}},\ }\href
  {\doibase 10.1103/PhysRevD.103.115023} {\bibfield  {journal} {\bibinfo
  {journal} {Phys. Rev. D}\ }\textbf {\bibinfo {volume} {103}},\ \bibinfo
  {pages} {115023} (\bibinfo {year} {2021}{\natexlab{b}})},\ \Eprint
  {http://arxiv.org/abs/2104.06417} {arXiv:2104.06417 [hep-ph]} \BibitemShut
  {NoStop}%
\bibitem [{\citenamefont {Dor\v{s}ner}\ \emph {et~al.}(2016)\citenamefont
  {Dor\v{s}ner}, \citenamefont {Fajfer}, \citenamefont {Greljo}, \citenamefont
  {Kamenik},\ and\ \citenamefont {Ko\v{s}nik}}]{Dorsner:2016wpm}%
  \BibitemOpen
  \bibfield  {author} {\bibinfo {author} {\bibfnamefont {I.}~\bibnamefont
  {Dor\v{s}ner}}, \bibinfo {author} {\bibfnamefont {S.}~\bibnamefont {Fajfer}},
  \bibinfo {author} {\bibfnamefont {A.}~\bibnamefont {Greljo}}, \bibinfo
  {author} {\bibfnamefont {J.~F.}\ \bibnamefont {Kamenik}}, \ and\ \bibinfo
  {author} {\bibfnamefont {N.}~\bibnamefont {Ko\v{s}nik}},\ }\href {\doibase
  10.1016/j.physrep.2016.06.001} {\bibfield  {journal} {\bibinfo  {journal}
  {Phys. Rept.}\ }\textbf {\bibinfo {volume} {641}},\ \bibinfo {pages} {1}
  (\bibinfo {year} {2016})},\ \Eprint {http://arxiv.org/abs/1603.04993}
  {arXiv:1603.04993 [hep-ph]} \BibitemShut {NoStop}%
\bibitem [{\citenamefont {Arnold}\ \emph {et~al.}(2013)\citenamefont {Arnold},
  \citenamefont {Fornal},\ and\ \citenamefont {Wise}}]{Arnold:2012sd}%
  \BibitemOpen
  \bibfield  {author} {\bibinfo {author} {\bibfnamefont {J.~M.}\ \bibnamefont
  {Arnold}}, \bibinfo {author} {\bibfnamefont {B.}~\bibnamefont {Fornal}}, \
  and\ \bibinfo {author} {\bibfnamefont {M.~B.}\ \bibnamefont {Wise}},\ }\href
  {\doibase 10.1103/PhysRevD.87.075004} {\bibfield  {journal} {\bibinfo
  {journal} {Phys. Rev. D}\ }\textbf {\bibinfo {volume} {87}},\ \bibinfo
  {pages} {075004} (\bibinfo {year} {2013})},\ \Eprint
  {http://arxiv.org/abs/1212.4556} {arXiv:1212.4556 [hep-ph]} \BibitemShut
  {NoStop}%
\bibitem [{\citenamefont {Crivellin}\ and\ \citenamefont
  {Schnell}(2022)}]{Crivellin:2021ejk}%
  \BibitemOpen
  \bibfield  {author} {\bibinfo {author} {\bibfnamefont {A.}~\bibnamefont
  {Crivellin}}\ and\ \bibinfo {author} {\bibfnamefont {L.}~\bibnamefont
  {Schnell}},\ }\href {\doibase 10.1016/j.cpc.2021.108188} {\bibfield
  {journal} {\bibinfo  {journal} {Comput. Phys. Commun.}\ }\textbf {\bibinfo
  {volume} {271}},\ \bibinfo {pages} {108188} (\bibinfo {year} {2022})},\
  \Eprint {http://arxiv.org/abs/2105.04844} {arXiv:2105.04844 [hep-ph]}
  \BibitemShut {NoStop}%
\bibitem [{\citenamefont {Murgui}\ and\ \citenamefont
  {Wise}(2021)}]{Murgui:2021bdy}%
  \BibitemOpen
  \bibfield  {author} {\bibinfo {author} {\bibfnamefont {C.}~\bibnamefont
  {Murgui}}\ and\ \bibinfo {author} {\bibfnamefont {M.~B.}\ \bibnamefont
  {Wise}},\ }\href {\doibase 10.1103/PhysRevD.104.035017} {\bibfield  {journal}
  {\bibinfo  {journal} {Phys. Rev. D}\ }\textbf {\bibinfo {volume} {104}},\
  \bibinfo {pages} {035017} (\bibinfo {year} {2021})},\ \Eprint
  {http://arxiv.org/abs/2105.14029} {arXiv:2105.14029 [hep-ph]} \BibitemShut
  {NoStop}%
\bibitem [{\citenamefont {Dor\v{s}ner}\ \emph
  {et~al.}(2022{\natexlab{a}})\citenamefont {Dor\v{s}ner}, \citenamefont
  {Fajfer},\ and\ \citenamefont {Sumensari}}]{Dorsner:2022twk}%
  \BibitemOpen
  \bibfield  {author} {\bibinfo {author} {\bibfnamefont {I.}~\bibnamefont
  {Dor\v{s}ner}}, \bibinfo {author} {\bibfnamefont {S.}~\bibnamefont {Fajfer}},
  \ and\ \bibinfo {author} {\bibfnamefont {O.}~\bibnamefont {Sumensari}},\
  }\href {\doibase 10.1007/JHEP05(2022)183} {\bibfield  {journal} {\bibinfo
  {journal} {JHEP}\ }\textbf {\bibinfo {volume} {05}},\ \bibinfo {pages} {183}
  (\bibinfo {year} {2022}{\natexlab{a}})},\ \Eprint
  {http://arxiv.org/abs/2202.08287} {arXiv:2202.08287 [hep-ph]} \BibitemShut
  {NoStop}%
\bibitem [{\citenamefont {Davidson}\ and\ \citenamefont
  {Descotes-Genon}(2010)}]{Davidson:2010uu}%
  \BibitemOpen
  \bibfield  {author} {\bibinfo {author} {\bibfnamefont {S.}~\bibnamefont
  {Davidson}}\ and\ \bibinfo {author} {\bibfnamefont {S.}~\bibnamefont
  {Descotes-Genon}},\ }\href {\doibase 10.1007/JHEP11(2010)073} {\bibfield
  {journal} {\bibinfo  {journal} {JHEP}\ }\textbf {\bibinfo {volume} {11}},\
  \bibinfo {pages} {073} (\bibinfo {year} {2010})},\ \Eprint
  {http://arxiv.org/abs/1009.1998} {arXiv:1009.1998 [hep-ph]} \BibitemShut
  {NoStop}%
\bibitem [{\citenamefont {Erler}\ and\ \citenamefont
  {Su}(2013)}]{Erler:2013xha}%
  \BibitemOpen
  \bibfield  {author} {\bibinfo {author} {\bibfnamefont {J.}~\bibnamefont
  {Erler}}\ and\ \bibinfo {author} {\bibfnamefont {S.}~\bibnamefont {Su}},\
  }\href {\doibase 10.1016/j.ppnp.2013.03.004} {\bibfield  {journal} {\bibinfo
  {journal} {Prog. Part. Nucl. Phys.}\ }\textbf {\bibinfo {volume} {71}},\
  \bibinfo {pages} {119} (\bibinfo {year} {2013})},\ \Eprint
  {http://arxiv.org/abs/1303.5522} {arXiv:1303.5522 [hep-ph]} \BibitemShut
  {NoStop}%
\bibitem [{\citenamefont {Helm}(1956)}]{PhysRev.104.1466}%
  \BibitemOpen
  \bibfield  {author} {\bibinfo {author} {\bibfnamefont {R.~H.}\ \bibnamefont
  {Helm}},\ }\href {\doibase 10.1103/PhysRev.104.1466} {\bibfield  {journal}
  {\bibinfo  {journal} {Phys. Rev.}\ }\textbf {\bibinfo {volume} {104}},\
  \bibinfo {pages} {1466} (\bibinfo {year} {1956})}\BibitemShut {NoStop}%
\bibitem [{\citenamefont {Hoferichter}\ \emph {et~al.}(2020)\citenamefont
  {Hoferichter}, \citenamefont {Men\'endez},\ and\ \citenamefont
  {Schwenk}}]{Hoferichter:2020osn}%
  \BibitemOpen
  \bibfield  {author} {\bibinfo {author} {\bibfnamefont {M.}~\bibnamefont
  {Hoferichter}}, \bibinfo {author} {\bibfnamefont {J.}~\bibnamefont
  {Men\'endez}}, \ and\ \bibinfo {author} {\bibfnamefont {A.}~\bibnamefont
  {Schwenk}},\ }\href {\doibase 10.1103/PhysRevD.102.074018} {\bibfield
  {journal} {\bibinfo  {journal} {Phys. Rev. D}\ }\textbf {\bibinfo {volume}
  {102}},\ \bibinfo {pages} {074018} (\bibinfo {year} {2020})},\ \Eprint
  {http://arxiv.org/abs/2007.08529} {arXiv:2007.08529 [hep-ph]} \BibitemShut
  {NoStop}%
\bibitem [{\citenamefont {Fricke}\ \emph {et~al.}(1995)\citenamefont {Fricke},
  \citenamefont {Bernhardt}, \citenamefont {Heilig}, \citenamefont {Schaller},
  \citenamefont {Schellenberg}, \citenamefont {Shera},\ and\ \citenamefont
  {de~Jager}}]{Fricke:1995zz}%
  \BibitemOpen
  \bibfield  {author} {\bibinfo {author} {\bibfnamefont {G.}~\bibnamefont
  {Fricke}}, \bibinfo {author} {\bibfnamefont {C.}~\bibnamefont {Bernhardt}},
  \bibinfo {author} {\bibfnamefont {K.}~\bibnamefont {Heilig}}, \bibinfo
  {author} {\bibfnamefont {L.~A.}\ \bibnamefont {Schaller}}, \bibinfo {author}
  {\bibfnamefont {L.}~\bibnamefont {Schellenberg}}, \bibinfo {author}
  {\bibfnamefont {E.~B.}\ \bibnamefont {Shera}}, \ and\ \bibinfo {author}
  {\bibfnamefont {C.~W.}\ \bibnamefont {de~Jager}},\ }\href {\doibase
  10.1006/adnd.1995.1007} {\bibfield  {journal} {\bibinfo  {journal} {Atom.
  Data Nucl. Data Tabl.}\ }\textbf {\bibinfo {volume} {60}},\ \bibinfo {pages}
  {177} (\bibinfo {year} {1995})}\BibitemShut {NoStop}%
\bibitem [{\citenamefont {Angeli}\ and\ \citenamefont
  {Marinova}(2013)}]{Angeli:2013epw}%
  \BibitemOpen
  \bibfield  {author} {\bibinfo {author} {\bibfnamefont {I.}~\bibnamefont
  {Angeli}}\ and\ \bibinfo {author} {\bibfnamefont {K.~P.}\ \bibnamefont
  {Marinova}},\ }\href {\doibase 10.1016/j.adt.2011.12.006} {\bibfield
  {journal} {\bibinfo  {journal} {Atom. Data Nucl. Data Tabl.}\ }\textbf
  {\bibinfo {volume} {99}},\ \bibinfo {pages} {69} (\bibinfo {year}
  {2013})}\BibitemShut {NoStop}%
\bibitem [{\citenamefont {Bender}\ \emph {et~al.}(1999)\citenamefont {Bender},
  \citenamefont {Rutz}, \citenamefont {Reinhard}, \citenamefont {Maruhn},\ and\
  \citenamefont {Greiner}}]{Bender:1999yt}%
  \BibitemOpen
  \bibfield  {author} {\bibinfo {author} {\bibfnamefont {M.}~\bibnamefont
  {Bender}}, \bibinfo {author} {\bibfnamefont {K.}~\bibnamefont {Rutz}},
  \bibinfo {author} {\bibfnamefont {P.~G.}\ \bibnamefont {Reinhard}}, \bibinfo
  {author} {\bibfnamefont {J.~A.}\ \bibnamefont {Maruhn}}, \ and\ \bibinfo
  {author} {\bibfnamefont {W.}~\bibnamefont {Greiner}},\ }\href {\doibase
  10.1103/PhysRevC.60.034304} {\bibfield  {journal} {\bibinfo  {journal} {Phys.
  Rev. C}\ }\textbf {\bibinfo {volume} {60}},\ \bibinfo {pages} {034304}
  (\bibinfo {year} {1999})},\ \Eprint {http://arxiv.org/abs/nucl-th/9906030}
  {arXiv:nucl-th/9906030} \BibitemShut {NoStop}%
\bibitem [{\citenamefont {Reinhard}\ and\ \citenamefont
  {Flocard}(1995)}]{Reinhard:1995zz}%
  \BibitemOpen
  \bibfield  {author} {\bibinfo {author} {\bibfnamefont {P.~G.}\ \bibnamefont
  {Reinhard}}\ and\ \bibinfo {author} {\bibfnamefont {H.}~\bibnamefont
  {Flocard}},\ }\href {\doibase 10.1016/0375-9474(94)00770-N} {\bibfield
  {journal} {\bibinfo  {journal} {Nucl. Phys. A}\ }\textbf {\bibinfo {volume}
  {584}},\ \bibinfo {pages} {467} (\bibinfo {year} {1995})}\BibitemShut
  {NoStop}%
\bibitem [{\citenamefont {Niksic}\ \emph {et~al.}(2008)\citenamefont {Niksic},
  \citenamefont {Vretenar},\ and\ \citenamefont {Ring}}]{Niksic:2008vp}%
  \BibitemOpen
  \bibfield  {author} {\bibinfo {author} {\bibfnamefont {T.}~\bibnamefont
  {Niksic}}, \bibinfo {author} {\bibfnamefont {D.}~\bibnamefont {Vretenar}}, \
  and\ \bibinfo {author} {\bibfnamefont {P.}~\bibnamefont {Ring}},\ }\href
  {\doibase 10.1103/PhysRevC.78.034318} {\bibfield  {journal} {\bibinfo
  {journal} {Phys. Rev. C}\ }\textbf {\bibinfo {volume} {78}},\ \bibinfo
  {pages} {034318} (\bibinfo {year} {2008})},\ \Eprint
  {http://arxiv.org/abs/0809.1375} {arXiv:0809.1375 [nucl-th]} \BibitemShut
  {NoStop}%
\bibitem [{\citenamefont {Payne}\ \emph {et~al.}(2019)\citenamefont {Payne},
  \citenamefont {Bacca}, \citenamefont {Hagen}, \citenamefont {Jiang},\ and\
  \citenamefont {Papenbrock}}]{Payne:2019wvy}%
  \BibitemOpen
  \bibfield  {author} {\bibinfo {author} {\bibfnamefont {C.~G.}\ \bibnamefont
  {Payne}}, \bibinfo {author} {\bibfnamefont {S.}~\bibnamefont {Bacca}},
  \bibinfo {author} {\bibfnamefont {G.}~\bibnamefont {Hagen}}, \bibinfo
  {author} {\bibfnamefont {W.}~\bibnamefont {Jiang}}, \ and\ \bibinfo {author}
  {\bibfnamefont {T.}~\bibnamefont {Papenbrock}},\ }\href {\doibase
  10.1103/PhysRevC.100.061304} {\bibfield  {journal} {\bibinfo  {journal}
  {Phys. Rev. C}\ }\textbf {\bibinfo {volume} {100}},\ \bibinfo {pages}
  {061304} (\bibinfo {year} {2019})},\ \Eprint
  {http://arxiv.org/abs/1908.09739} {arXiv:1908.09739 [nucl-th]} \BibitemShut
  {NoStop}%
\bibitem [{\citenamefont {Collar}\ \emph {et~al.}(2019)\citenamefont {Collar},
  \citenamefont {Kavner},\ and\ \citenamefont {Lewis}}]{Collar:2019ihs}%
  \BibitemOpen
  \bibfield  {author} {\bibinfo {author} {\bibfnamefont {J.~I.}\ \bibnamefont
  {Collar}}, \bibinfo {author} {\bibfnamefont {A.~R.~L.}\ \bibnamefont
  {Kavner}}, \ and\ \bibinfo {author} {\bibfnamefont {C.~M.}\ \bibnamefont
  {Lewis}},\ }\href {\doibase 10.1103/PhysRevD.100.033003} {\bibfield
  {journal} {\bibinfo  {journal} {Phys. Rev. D}\ }\textbf {\bibinfo {volume}
  {100}},\ \bibinfo {pages} {033003} (\bibinfo {year} {2019})},\ \Eprint
  {http://arxiv.org/abs/1907.04828} {arXiv:1907.04828 [nucl-ex]} \BibitemShut
  {NoStop}%
\bibitem [{\citenamefont {Akimov}\ \emph {et~al.}(2018)\citenamefont {Akimov}
  \emph {et~al.}}]{COHERENT:2018imc}%
  \BibitemOpen
  \bibfield  {author} {\bibinfo {author} {\bibfnamefont {D.}~\bibnamefont
  {Akimov}} \emph {et~al.} (\bibinfo {collaboration} {COHERENT}),\ }\href
  {\doibase 10.5281/zenodo.1228631} {\  (\bibinfo {year} {2018}),\
  10.5281/zenodo.1228631},\ \Eprint {http://arxiv.org/abs/1804.09459}
  {arXiv:1804.09459 [nucl-ex]} \BibitemShut {NoStop}%
\bibitem [{\citenamefont {Tumasyan}\ \emph {et~al.}(2021)\citenamefont
  {Tumasyan} \emph {et~al.}}]{CMS:2021far}%
  \BibitemOpen
  \bibfield  {author} {\bibinfo {author} {\bibfnamefont {A.}~\bibnamefont
  {Tumasyan}} \emph {et~al.} (\bibinfo {collaboration} {CMS}),\ }\href
  {\doibase 10.1007/JHEP11(2021)153} {\bibfield  {journal} {\bibinfo  {journal}
  {JHEP}\ }\textbf {\bibinfo {volume} {11}},\ \bibinfo {pages} {153} (\bibinfo
  {year} {2021})},\ \Eprint {http://arxiv.org/abs/2107.13021} {arXiv:2107.13021
  [hep-ex]} \BibitemShut {NoStop}%
\bibitem [{\citenamefont {Taxil}\ \emph {et~al.}(2000)\citenamefont {Taxil},
  \citenamefont {Tugcu},\ and\ \citenamefont {Virey}}]{Taxil:1999pf}%
  \BibitemOpen
  \bibfield  {author} {\bibinfo {author} {\bibfnamefont {P.}~\bibnamefont
  {Taxil}}, \bibinfo {author} {\bibfnamefont {E.}~\bibnamefont {Tugcu}}, \ and\
  \bibinfo {author} {\bibfnamefont {J.~M.}\ \bibnamefont {Virey}},\ }\href
  {\doibase 10.1007/s100520050743} {\bibfield  {journal} {\bibinfo  {journal}
  {Eur. Phys. J. C}\ }\textbf {\bibinfo {volume} {14}},\ \bibinfo {pages} {165}
  (\bibinfo {year} {2000})},\ \Eprint {http://arxiv.org/abs/hep-ph/9912272}
  {arXiv:hep-ph/9912272} \BibitemShut {NoStop}%
\bibitem [{\citenamefont {Dor\v{s}ner}\ \emph
  {et~al.}(2022{\natexlab{b}})\citenamefont {Dor\v{s}ner}, \citenamefont
  {Lejli\'c},\ and\ \citenamefont {Saad}}]{Dorsner:2022ibm}%
  \BibitemOpen
  \bibfield  {author} {\bibinfo {author} {\bibfnamefont {I.}~\bibnamefont
  {Dor\v{s}ner}}, \bibinfo {author} {\bibfnamefont {A.}~\bibnamefont
  {Lejli\'c}}, \ and\ \bibinfo {author} {\bibfnamefont {S.}~\bibnamefont
  {Saad}},\ }\href@noop {} {\  (\bibinfo {year} {2022}{\natexlab{b}})},\
  \Eprint {http://arxiv.org/abs/2210.11004} {arXiv:2210.11004 [hep-ph]}
  \BibitemShut {NoStop}%
\bibitem [{\citenamefont {Dey}\ \emph {et~al.}(2018)\citenamefont {Dey},
  \citenamefont {Kar}, \citenamefont {Mitra}, \citenamefont {Spannowsky},\ and\
  \citenamefont {Vincent}}]{Dey:2017ede}%
  \BibitemOpen
  \bibfield  {author} {\bibinfo {author} {\bibfnamefont {U.~K.}\ \bibnamefont
  {Dey}}, \bibinfo {author} {\bibfnamefont {D.}~\bibnamefont {Kar}}, \bibinfo
  {author} {\bibfnamefont {M.}~\bibnamefont {Mitra}}, \bibinfo {author}
  {\bibfnamefont {M.}~\bibnamefont {Spannowsky}}, \ and\ \bibinfo {author}
  {\bibfnamefont {A.~C.}\ \bibnamefont {Vincent}},\ }\href {\doibase
  10.1103/PhysRevD.98.035014} {\bibfield  {journal} {\bibinfo  {journal} {Phys.
  Rev. D}\ }\textbf {\bibinfo {volume} {98}},\ \bibinfo {pages} {035014}
  (\bibinfo {year} {2018})},\ \Eprint {http://arxiv.org/abs/1709.02009}
  {arXiv:1709.02009 [hep-ph]} \BibitemShut {NoStop}%
\bibitem [{\citenamefont {Be\v{c}irevi\'c}\ \emph {et~al.}(2018)\citenamefont
  {Be\v{c}irevi\'c}, \citenamefont {Panes}, \citenamefont {Sumensari},\ and\
  \citenamefont {Zukanovich~Funchal}}]{Becirevic:2018uab}%
  \BibitemOpen
  \bibfield  {author} {\bibinfo {author} {\bibfnamefont {D.}~\bibnamefont
  {Be\v{c}irevi\'c}}, \bibinfo {author} {\bibfnamefont {B.}~\bibnamefont
  {Panes}}, \bibinfo {author} {\bibfnamefont {O.}~\bibnamefont {Sumensari}}, \
  and\ \bibinfo {author} {\bibfnamefont {R.}~\bibnamefont
  {Zukanovich~Funchal}},\ }\href {\doibase 10.1007/JHEP06(2018)032} {\bibfield
  {journal} {\bibinfo  {journal} {JHEP}\ }\textbf {\bibinfo {volume} {06}},\
  \bibinfo {pages} {032} (\bibinfo {year} {2018})},\ \Eprint
  {http://arxiv.org/abs/1803.10112} {arXiv:1803.10112 [hep-ph]} \BibitemShut
  {NoStop}%
\bibitem [{\citenamefont {Abbiendi}\ \emph {et~al.}(2013)\citenamefont
  {Abbiendi} \emph {et~al.}}]{ALEPH:2013htx}%
  \BibitemOpen
  \bibfield  {author} {\bibinfo {author} {\bibfnamefont {G.}~\bibnamefont
  {Abbiendi}} \emph {et~al.} (\bibinfo {collaboration} {ALEPH, DELPHI, L3,
  OPAL, LEP}),\ }\href {\doibase 10.1140/epjc/s10052-013-2463-1} {\bibfield
  {journal} {\bibinfo  {journal} {Eur. Phys. J. C}\ }\textbf {\bibinfo {volume}
  {73}},\ \bibinfo {pages} {2463} (\bibinfo {year} {2013})},\ \Eprint
  {http://arxiv.org/abs/1301.6065} {arXiv:1301.6065 [hep-ex]} \BibitemShut
  {NoStop}%
\end{thebibliography}%

\end{document}